\documentclass[12pt]{article}
\usepackage{amsfonts}
\usepackage{amssymb}
\usepackage{amsmath}
\usepackage{geometry}
\usepackage{color}
\usepackage{graphicx}% Include figure files

\input{tcilatex}

\begin{document}

\title{\textcolor{blue}{Divergence of logarithm of a unimodular monodromy matrix near the
edges of the Brillouin zone}}
\author{A. L. Shuvalov$^{1}$, A. A. Kutsenko$^{2}$, A. N. Norris$^{3}$ \\ \\
%EndAName
$^{1}$ Laboratoire de M\'{e}canique Physique, UMR CNRS 5469, \\
Universit\'{e} Bordeaux 1, Talence 33405, France\\
$^{2}\ $Mathematics \& Mechanics Faculty, St-Petersburg State University, \\
St. Petersburg 198504 , Russia\\
$^{3}\ $Mechanical and Aerospace Engineering, Rutgers University,\\
Piscataway NJ 08854-8058, USA}
\maketitle

\begin{abstract}
A first-order ordinary differential system with a matrix of periodic
coefficients $\mathbf{Q}\left( y\right) =\mathbf{Q}\left( y+T\right) $ is
studied in the context of time-harmonic elastic waves travelling with
frequency $\omega $ in a unidirectionally periodic medium, for which case
the monodromy matrix $\mathbf{M}\left( \omega \right) $ implies a propagator
of the wave field over a period. The main interest to the matrix logarithm $%
\ln \mathbf{M}\left( \omega \right) $ is owing to the fact that it yields
the 'effective' matrix $\mathbf{Q}_{\mathrm{eff}}\left( \omega \right) $ of
the dynamic-homogenization method. For the typical case of a unimodular
matrix $\mathbf{M}\left( \omega \right) $ ($\det \mathbf{M}=1$), it is
established that the components of $\ln \mathbf{M}\left( \omega \right) $
diverge as $\left( \omega -\omega _{0}\right) ^{-1/2}$ with $\omega
\rightarrow \omega _{0},$ where $\omega _{0}$ is the set of frequencies of
the passband/stopband crossovers at the edges of the first Brillouin zone.
The divergence disappears for a homogeneous medium. Mathematical and
physical aspects of this observation are discussed. Explicit analytical
examples of $\mathbf{Q}_{\mathrm{eff}}\left( \omega \right) $ and of its
diverging asymptotics at $\omega \rightarrow \omega _{0}$ are provided for a
simple model of scalar waves in a two-component periodic structure
consisting of identical bilayers or layers in spring-mass-spring contact.
The case of high contrast due to stiff/soft layers or soft springs is
elaborated. Special attention in this case is given to the asymptotics of $%
\mathbf{Q}_{\mathrm{eff}}\left( \omega \right) $ near the first stopband
that occurs at the Brillouin-zone edge at arbitrary low frequency. \medskip
The link to the quasi-static asymptotics of the same $\mathbf{Q}_{\mathrm{eff%
}}\left( \omega \right) $ near the point $\omega =0$ is also
elucidated.\medskip

\end{abstract}

\section{Introduction}

The first-order ordinary differential system 
\begin{equation}
\mathbf{Q}\left( y\right) \mathbf{\eta }\left( y\right) =\frac{\mathrm{d}}{%
\mathrm{d}y}\mathbf{\eta }\left( y\right)  \label{1}
\end{equation}%
with a $n\times n$ matrix of continuous or piecewise continuous periodic
coefficients $\mathbf{Q}\left( y\right) =\mathbf{Q}\left( y+T\right) $ is a
classical problem arising in miscellaneous models of applied mathematics and
mathematical physics. Its analysis largely relies on the Floquet theorem
asserting that the matricant $\mathbf{M}\left( y,0\right) $, which is the
fundamental solution of (\ref{1}) yielding $\mathbf{\eta }\left( y\right) =%
\mathbf{M}\left( y,0\right) \mathbf{\eta }\left( 0\right) $, can be factored
into the product%
\begin{equation}
\mathbf{M}\left( y,0\right) =\mathbf{L}\left( y\right) \exp \left( i\mathbf{K%
}y\right) ,  \label{2}
\end{equation}%
where $\mathbf{L}\left( y\right) =\mathbf{L}\left( y+T\right) $~($=\mathbf{M}%
\left( y_{n},0\right) \exp \left( -i\mathbf{K}y_{n}\right) $ with $%
y_{n}=y\left( \func{mod}nT\right) $),$\ \mathbf{L}\left( 0\right) =\mathbf{I}
$ ($\mathbf{I}$ is the identity matrix), and $\mathbf{K}$ is a constant
matrix \cite{Pease}. By (\ref{2}), $\mathbf{K}$ is defined by the equation 
\begin{equation}
\exp \left( i\mathbf{K}T\right) =\mathbf{M}\left( T,0\right) \
\Longrightarrow \ i\mathbf{K}T=\ln \mathbf{M}\left( T,0\right) ,  \label{3}
\end{equation}%
where $\mathbf{M}\left( T,0\right) \equiv \mathbf{M}$ is termed the
monodromy matrix (its reference to $\left( T,0\right) $ is dropped
hereafter). It can be calculated by a number of available methods, e.g.,
using the Peano series of multiple integrals of $\mathbf{Q}\left( y\right) $%
, or applying polynomial expansion of $\mathbf{Q}\left( y\right) ,$ or
discretizing $\mathbf{Q}\left( y\right) $. In many problems the system
matrix $\mathbf{Q}\left( y\right) $ is a continuous function of a certain
control parameter $\omega ,$ and hence also $\mathbf{M=M}\left( \omega
\right) .$ At first glance, the matrix logarithm $\mathbf{K}\left( \omega
\right) $ is well-behaved as long as the logarithm of the eigenvalues $%
q\left( \omega \right) $ of $\mathbf{M}\left( \omega \right) $ is
well-behaved. However, it turns out that $\ln \mathbf{M}\left( \omega
\right) $ diverges at $\omega \rightarrow \omega _{0},$ where $\omega _{0}$
corresponds to a non-semisimple (not diagonalizable) $\mathbf{M}\left(
\omega _{0}\right) $ with a degenerate eigenvalue $q\left( \omega
_{0}\right) $ whose values, being taken on the same Riemann sheet of $\ln q,$
are situated on the opposite edges of the cut. This fairly surprising
observation seems to have passed unnoticed in the extensive reference
literature on the matrix logarithm. The manner in which such a divergence
reveals itself in the Floquet formalism is discussed in the present paper in
the context where Eq. (\ref{1}) is associated with time-harmonic elastic
waves travelling at frequency $\omega $ in unidirectionally (1D) periodic
media. Within this context, the system (\ref{1}) such that consists of $n=2$
equations and hence is equivalent to Hill's equation describes scalar
acoustic (or electromagnetic) waves \cite{B,MW}; the cases where (\ref{1})
consists of $n=4,6,8...$ equations corresponds to coupled waves in elastic
isotropic or anisotropic media, in piezoelectric or piezomagnetoelectric
media, etc. In either of these cases, the monodromy matrix $\mathbf{M}$ is
often called the propagator (of the wave field) over the period $T.$

The matrix logarithm $\mathbf{K}\left( \omega \right) $ (\ref{3}) is a
crucial ingredient in the dynamic-homogenization approach. Assuming that $%
\exp \left( i\mathbf{K}y\right) $ in the 1D Floquet theorem (\ref{2}) is a
relatively slowly varying function, this approach seeks to replace an exact
solution $\mathbf{M}\left( y,0\right) $ by its 'slow component' $\exp \left(
i\mathbf{K}y\right) $ and hence to replace the actual periodically
inhomogeneous material by an 'homogenized' medium with spatially constant
but frequency dispersive properties described by the 'effective' matrix 
\begin{equation}
\mathbf{Q}_{\mathrm{eff}}\left( \omega \right) =i\mathbf{K}\left( \omega
\right) ,  \label{3.1}
\end{equation}%
see e.g. \cite{N,PBG,WR}. Obviously, the matrix $\mathbf{Q}_{\mathrm{eff}}$
also provides (regardless of any assumptions) an exact solution $\mathbf{M}%
\left( nT,0\right) =\exp \left( in\mathbf{K}T\right) $ at the interfaces
between the periods. Another aspect of the matrix logarithm $\mathbf{K}%
\left( \omega \right) $ is related to the Floquet dispersion branches $%
\omega \left( K\right) $ or $K\left( \omega \right) $. These are determined
by the secular equation for $\mathbf{M}$, 
\begin{equation}
\det \left[ \mathbf{M}\left( \omega \right) -q\left( \omega \right) \mathbf{I%
}\right] =0,  \label{6}
\end{equation}%
so that the definition $q=e^{iKT}$ yields $iK\left( \omega \right) T=\ln
q\left( \omega \right) ,$ or else by the formally equivalent secular
equation for $\mathbf{K,}$ 
\begin{equation}
\det \left[ \mathbf{K}\left( \omega \right) -K\left( \omega \right) \mathbf{I%
}\right] =0.  \label{5}
\end{equation}%
The Floquet spectrum is commonly defined over the first Brillouin zone (BZ) $%
\func{Re}KT\in \left[ -\pi ,\pi \right] ,$ which is related to the zeroth
Riemann sheet of the single-valued $\ln q=\ln \left\vert q\right\vert +i\arg
q$ with the cut $\arg q=\pm \pi $ corresponding to the BZ edges. The
frequency intervals where $K$ is real or complex are called passbands and
stopbands, respectively.

The paper is concerned with the typical case where $\mathbf{M}\left( \omega
\right) $ is unimodular ($\det \mathbf{M}=1$) and so the BZ edges contain
the passband/stopband crossovers at a set of frequencies $\omega =\omega
_{0} $ associated with a degenerate pair of eigenvalues $q\left( \omega
_{0}\right) $ of$\ \mathbf{M}\left( \omega _{0}\right) .$ According to the
background outlined in \S 2, this is the case for a normal propagation
across an arbitrary anisotropic periodic structure or for an arbitrary
propagation direction in the presence of appropriately oriented symmetry
plane. The original material of this work consists of two parts, \S\ 3 and 
\S 4. The first part (\S 3) deals with the problem in general. It is shown
that the matrix $\ln \mathbf{M}\left( \omega \right) =i\mathbf{K}\left(
\omega \right) T,$ and hence $\mathbf{Q}_{\mathrm{eff}},$ must have
components diverging as $\left( \omega -\omega _{0}\right) ^{-1/2}$ when $%
\omega \rightarrow \omega _{0},$ i.e. when the real Floquet branches tend to
the BZ edges or the complex part of $\pm K=\pi /T+i\func{Im}K$ tends to
zero. The eigenspectrum of $\mathbf{K}\left( \omega \right) $ certainly
remains well-behaved for any $\omega $ infinitesimally close to $\omega _{0}$%
; however, computing the Floquet spectrum $K\left( \omega \right) $
specifically from Eq. (\ref{5}) may become numerically unstable at $\omega $
close to $\omega _{0}.$ A transition is explained from a weakly
inhomogeneous to perfectly homogeneous elastic medium, for which $\ln 
\mathbf{M}\left( \omega \right) $ certainly does not diverge. The second
part (\S 4) presents detailed analytical examples of $\mathbf{Q}_{\mathrm{eff%
}}\left( \omega \right) $ and of its diverging asymptotics for $\omega
\rightarrow \omega _{0}$ for the shear-horizontal wave in a periodic
structure composed of piecewise homogeneous bilayers or layers in
spring-mass-spring contact. Particular attention is given to the
high-contrast case with either a soft layer in the bilayer or with a soft
spring in the interfacial joint. The interest to this case lies in the fact
that the first stopband at the BZ edges and hence the local divergence of $%
\mathbf{Q}_{\mathrm{eff}}\left( \omega \right) $ occurs at low frequency
that may in principle be made arbitrarily small. To this end, a link to the
regular asymptotics of the same $\mathbf{Q}_{\mathrm{eff}}\left( \omega
\right) $ near the point $\omega =0$ is also elucidated. The basic points of
the study are summarized in \S 5. Some technical aspects of the derivations
of \S 3 and \S 4 are detailed in the Appendix.

\section{Background}

Consider elastic waves in a 1D-periodic infinite anisotropic non-absorbing
medium without sources. Choose the periodicity direction as the axis $Y$ and
denote the (least) period by $T,$ so that the density and the elasticity
tensor satisfy $\rho \left( y\right) =\rho \left( y+T\right) $ and $\mathbf{c%
}\left( y\right) =\mathbf{c}\left( y+T\right) $, respectively. Take the axis 
$X$ in the sagittal plane spanned by $Y$ and by the direction to the
observation point. Applying Fourier transforms in time and in $X$ brings in
the frequency $\omega $ and wavenumber $k_{x}$ as the (real) parameters of
the problem.

The equation of motion and the linear stress-strain law may be combined into
the system (\ref{1}) of, generally, six equations. The periodic matrix of
coefficients $\mathbf{Q}\left( y\right) $ defined through $\rho \left(
y\right) ,$ $\mathbf{c}\left( y\right) $ and $\omega $, $k_{x},$ is pure
imaginary and has the Hamiltonian structure%
\begin{equation}
\mathbf{Q}\left( y\right) =\mathbf{TQ}^{\mathrm{T}}\left( y\right) \mathbf{T,%
}  \label{8}
\end{equation}%
where the superscript $^{\mathrm{T}}$ means transpose and $\mathbf{T}$ is
the matrix with zero diagonal and identity off-diagonal 3$\times $3 blocks
(see e.g. \cite{WaMot} for the details).

In the following we deal with the essentially typical case of a medium with
at least a single symmetry plane $m$ orthogonal to the axis $X$ or $Y.$ Then
the trace of $\mathbf{Q}\left( y\right) $ is zero for any $y$. Therefore, by
the Jacobi identity, $\mathbf{M}\left( y,0\right) $ is unimodular and hence
so is the monodromy matrix $\mathbf{M}\equiv \mathbf{M}\left( T,0\right) ,$
i.e. 
\begin{equation}
\det \mathbf{M}=1.  \label{9}
\end{equation}%
The identities (\ref{8}) and (\ref{9}) together ensure that\ for\ every\
eigenvalue\ $q_{\alpha }$\ of\ $\mathbf{M}$,\ there\ is\ a\ corresponding\
eigenvalue\ $q_{\beta }=1/q_{\alpha }$ where $\alpha ,\beta =1,...,6.$ This
property has been established in \cite{BH} for a piecewise constant $\mathbf{%
Q}\left( y\right) $ and $m\perp Y;$ its generalization for any piecewise
continuous $\mathbf{Q}\left( y\right) $ and for $m\perp X$ is obvious. Note
that no stipulation of any material symmetry is needed if the wave
propagates strictly along the periodicity direction $Y$ (i.e. if $k_{x}=0$),
which is when (\ref{9}) is always true. Also note that the out-of-plane
motion with respect to the symmetry plane $m\perp Z$ of a monoclinic body
(which has no other symmetry planes) can be cast in the form with property (%
\ref{9}), see \cite{WaMot1}.

Let $\omega $ be a single free dispersion parameter ($k_{x}$ is fixed or
expressed through $\omega $). Each pair $q_{\beta }\left( \omega \right)
=1/q_{\alpha }\left( \omega \right) $ corresponds to a set of dispersion
curves $K_{\alpha }\left( \omega \right) =-K_{\beta }\left( \omega \right) $
in the BZ $\func{Re}K_{\alpha ,\beta }T\in \left[ -\pi ,\pi \right] ,$ which
are symmetric about the line $K=0.$ In view of (\ref{9}), the eigenvalues $%
q=1$ and $q=-1,$ occurring, respectively, at the centre and edges of the BZ,
are assuredly degenerate. We are interested in the case $q=-1,$ which is
associated with a sequence of passband/stopband crossover points at the BZ
edges, and specifically in the behaviour of the matrix $\ln \mathbf{M}=i%
\mathbf{K}T$ in the vicinity of these points.

\section{Divergence of $\mathbf{K}\left( \protect\omega \right) $ near the
BZ edges}

\subsection{Derivation}

Denote by $\omega =\omega _{0}$ the frequency, at which some pair of
eigenvalue branches $q_{1}\left( \omega \right) =1/q_{2}\left( \omega
\right) $ of the monodromy matrix $\mathbf{M}\left( \omega \right) $ fall
into two-fold degeneracy $q_{1}\left( \omega _{0}\right) =q_{2}\left( \omega
_{0}\right) =-1$ rendering $\mathbf{M}\left( \omega _{0}\right) $
non-semisimple. Consider a function $\ln q=\ln \left\vert q\right\vert
+i\arg q$ defined on the zeroth Riemann sheet with a cut $\arg q=\pm \pi $
passing through $-1.$ Let $\omega $ lying in the stopband or passband tend
to $\omega _{0}$ from, respectively, above or below. Then $q_{1}\left(
\omega \right) $ and $q_{2}\left( \omega \right) $ tend to $e^{\pm i\pi },$
thus approaching their degenerate value $-1$ from the opposite sides of the
cut for $\ln q$, and, correspondingly, $\ln q_{1,2}\left( \omega \right)
=iK_{1,2}\left( \omega \right) T$ tend to $\pm i\pi ,$ meaning that two
Floquet branches tend to the opposite edges of the BZ.

This is indeed nothing else than a very standard setup. The state of affairs
is, however, not so trivial when the same limit $\omega \rightarrow \omega
_{0}$ is applied to the matrix logarithm $\ln \mathbf{M}\left( \omega
\right) =i\mathbf{K}\left( \omega \right) T.$ It is natural to specify it by
asking that both eigenvalues $\ln q_{1,2}\left( \omega \right) $ of $\ln 
\mathbf{M}\left( \omega \right) $ satisfy the above-mentioned definition of $%
\ln q$ (the issue of alternative definitions of $\ln \mathbf{M}$ is
addressed in \S 3.1 and in \S A.2 of Appendix). As we have just observed,
these eigenvalues tend to $\pm i\pi $ as $\omega \rightarrow \omega _{0},$
i.e. they do not approach each other in contrast to the eigenvalues $%
q_{1}\left( \omega \right) \rightarrow q_{2}\left( \omega \right) $ of $%
\mathbf{M}\left( \omega \right) $. This signals a singularity of $\ln 
\mathbf{M}\left( \omega \right) $ on the path $\omega \rightarrow \omega
_{0}.$

Let us analyze the local behaviour of $\ln \mathbf{M}\left( \omega \right) $
for $\omega =\omega _{0}+\Delta \omega $ ($\left\vert \Delta \omega /\omega
_{0}\right\vert \ll 1$). With reference to (\ref{9}), denote 
\begin{equation}
q_{1,2}\left( \omega _{0}+\Delta \omega \right) \approx q_{d}\pm \delta q%
\underset{\omega \rightarrow \omega _{0}}{\rightarrow }q_{1,2}\left( \omega
_{0}\right) \equiv q_{d}=-1,  \label{10}
\end{equation}%
where $\delta q$\ means the leading-order correction in the small parameter%
\textbf{\ }$\Delta \omega /\omega _{0}$. For brevity, assume the case of 2$%
\times $2 matrices (the same derivation for the general $n\times n$ case is
detailed in Appendix, \S\ A1). A polynomial formula for a function of a 2$%
\times $2 matrix $\mathbf{M}$ with eigenvalues $q_{1}\neq q_{2}$ has a
simple form 
\begin{equation}
f\left( \mathbf{M}\right) =\frac{q_{2}f\left( q_{1}\right) -q_{1}f\left(
q_{2}\right) }{q_{2}-q_{1}}\mathbf{I}+\frac{f\left( q_{2}\right) -f\left(
q_{1}\right) }{q_{2}-q_{1}}\mathbf{M,\ }  \label{11}
\end{equation}%
see e.g. \cite{HJ}. Taking (\ref{11}) for $\mathbf{M}\left( \omega
_{0}+\Delta \omega \right) =\mathbf{M}\left( \omega _{0}\right) +\mathbf{%
\Delta M}$ with $q_{1,2}\left( \omega \right) $ given by (\ref{10}) yields 
\begin{equation}
f\left[ \mathbf{M}\left( \omega _{0}+\Delta \omega \right) \right] =\frac{%
f_{01}+f_{02}}{2}\mathbf{I}+\left[ \frac{f_{01}-f_{02}}{2\delta q}+f^{\prime
}\left( q_{d}\right) \right] \left[ \mathbf{M}\left( \omega _{0}\right) +%
\mathbf{\Delta M}-q_{d}\mathbf{I}\right] +\mathbf{O}\left( \delta q\right) ,
\label{12}
\end{equation}%
where $f_{01,02}=\lim_{\omega \rightarrow \omega _{0}}f\left( q_{1,2}\left(
\omega \right) \right) $ and $\mathbf{O}$ is a matrix symbol 'of the order
of'. For the case in hand $f=\ln $ and $f_{01,02}=\ln \left( e^{\pm i\pi
}\right) ,$ whence (\ref{12}) becomes%
\begin{equation}
\ln \mathbf{M}\left( \omega _{0}+\Delta \omega \right) =\left( \frac{i\pi }{%
\delta q}+\frac{1}{q_{d}}\right) \left[ \mathbf{M}\left( \omega _{0}\right)
-q_{d}\mathbf{I}\right] +\frac{i\pi }{\delta q}\mathbf{\Delta M}+\mathbf{O}%
\left( \delta q,\mathbf{\Delta M}\right) .  \label{13}
\end{equation}%
Since $\mathbf{M}\left( \omega _{0}\right) -q_{d}\mathbf{I}$ is non-zero for
a non-semisimple $\mathbf{M}\left( \omega _{0}\right) $ while $\delta q$
tends to zero with $\Delta \omega \rightarrow 0,$ we conclude from Eq. (\ref%
{13}) that the matrix logarithm $\ln \mathbf{M}\left( \omega \right) ,$ and
thus $\mathbf{K}\left( \omega \right) ,$ must have components tending to
infinity when $\omega \rightarrow \omega _{0}.$ Note in passing that an
identically zero determinant of the first matrix term on the right-hand side
of (\ref{13}) does certainly not preclude but, on the contrary, underlies
(with due regard for the next term) the necessary identity $\det \left[ \ln 
\mathbf{M}\left( \omega \right) \right] \rightarrow \pi ^{2}$ as $\omega
\rightarrow \omega _{0}.$

Let us now find an asymptotic rate of divergence of $\ln \mathbf{M}\left(
\omega \right) $ in terms of $\Delta \omega \left( \equiv \omega -\omega
_{0}\right) .$ For a non-semisimple $\mathbf{M}$ of 2$\times $2 dimension,
the leading-order dependence $\delta q\sim \left( \omega -\omega _{0}\right)
^{1/2}$ obviously follows from a quadratic secular equation (\ref{6}). For
the general $n\times n$ case\textbf{,} the same trend is easy to infer from
the leading-order Taylor expansion of $D\left( q,\omega \right) \equiv \det %
\left[ \mathbf{M}\left( \omega \right) -q\mathbf{I}\right] $ about the point
of double degeneracy $q_{1,2}\left( \omega _{0}\right) =q_{d},$ which leads
to%
\begin{equation}
\left( \delta q\right) ^{2}=B\Delta \omega ,\ B=-2\left( \frac{\partial
D/\partial \omega }{\partial ^{2}D/\partial q^{2}}\right) _{\omega
_{0},q_{d}}.  \label{14}
\end{equation}%
Omitting details (see e.g. \cite{WaMot2}), it suffices to note that $B$ is
generally non-zero for non-semisimple $\mathbf{M}\left( \omega _{0}\right) .$
Thus, by (\ref{13}), $\ln \mathbf{M}\left( \omega \right) ,$ and hence $%
\mathbf{K}\left( \omega \right) ,$ diverges as $\left( \omega -\omega
_{0}\right) ^{-1/2}$ with $\omega \rightarrow \omega _{0}.$ An explicit form
of the coefficient $B$ will be exemplified in \S 4.

\subsection{Discussion}

A few formal remarks are in order. First it is reiterated that even though
the components of the dealt-with matrix $\ln \mathbf{M}\left( \omega \right)
=i\mathbf{K}\left( \omega \right) T$ diverge as $\omega \rightarrow \omega
_{0},$ its eigenvalues $\ln q_{1,2}=iK_{1,2}\left( \omega \right) T$ remain
formally well-defined so long as $\omega \neq \omega _{0}.$ It is also
understood that the exponential of this $\ln \mathbf{M}\left( \omega \right) 
$ at any $\omega \neq \omega _{0}$ certainly reproduces (continuous) $%
\mathbf{M}\left( \omega \right) $. Regarding the infinity of $\ln \mathbf{M}%
\left( \omega \right) $ precisely at $\omega =\omega _{0},$ which is when $%
\delta q=0$ on the right-hand side of (\ref{13}), it simply tells us that
the conventional definition of $\ln \mathbf{M}\left( \omega \right) ,$ which
refers both eigenvalues $\ln q_{1,2}\left( \omega \right) $ to the zeroth
Riemann sheet of $\ln q$ with the cut $\arg q=\pm \pi $ fixing the edges of
the BZ $\func{Re}KT\in \left[ -\pi ,\pi \right] ,$ precludes this matrix
function of $\omega $ from reaching the limiting point $\omega _{0}$ of the
path $\omega \rightarrow \omega _{0}$ continuously.

It is clear from the above that shifting the cut in the $q$-plane away from
the point $q=-1$ while keeping $\ln q_{1,2}$ on the same Riemann sheet leads
to a different matrix logarithm $\ln \mathbf{M}\left( \omega \right) $\ that
has degenerate eigenvalues $\ln q_{1}\left( \omega _{0}\right) =\ln
q_{2}\left( \omega _{0}\right) $\ and hence is well-behaved at $\omega
=\omega _{0}$\ and around it. However, this 'gain' for $\omega $ near $%
\omega _{0}$ is at the expense of one or another essential deficiency
elsewhere for the redefined $\ln \mathbf{M}\left( \omega \right) .$\ For
instance, if the eigenvalues $\ln q_{1,2}$ of $\ln \mathbf{M}\left( \omega
\right) $ are taken on the zeroth Riemann sheet with the cut $\arg q=0,2\pi
, $ then this $\ln \mathbf{M}\left( \omega \right) $ has the same divergence 
$\sim \left( \omega -\omega _{01}\right) ^{-1/2}$ due to the degeneracy $%
q_{1,2}\left( \omega _{01}\right) =1$ at the set $\omega _{01}$ of
passband/stopband crossovers occurring at $K=0,2\pi .$ An exception is the
origin point $\omega =0,$ where $\mathbf{M=I}$ and so any $\ln \mathbf{M}$
is continuous; however, the low-frequency onset of $\ln \mathbf{M}$ defined
by taking the cut $\arg q=0,2\pi $ has no physical sense (see Appendix, \S %
A2). Another possibility is to use a cut $\arg q=\varphi ,\varphi -2\pi $ at 
$\varphi \neq \pi n,$ e.g., at $\varphi $ such that $0<\varphi <\pi .$ Then $%
\ln \mathbf{M,}$ whose eigenvalues $\ln q_{1,2}=\pm iKT$ lie on the zeroth
Riemann sheet, is well-behaved with $\left\vert \arg q\right\vert
=\left\vert KT\right\vert $ growing from zero but only until reaching $%
\varphi ,$ where there is a jump to a different matrix $\ln \mathbf{M,}$ for
which the eigenvalue $\ln q_{1}$ has to be shifted from $\arg q_{1}=KT$ to $%
\mathrm{\arg }q_{1}=KT-2\pi $ with $KT>0$ increasing above $\varphi $. Note
that a similar piecewise discontinuity pertains in the BZ $\func{Re}KT\in %
\left[ -\pi ,\pi \right] $ to the logarithm of $\mathbf{M}$ that is not
unimodular ($\det \mathbf{M}\neq 1$)$.$ Thus, using any 'unconventional'
definition of the logarithm of $\mathbf{M}$ based on shifting the cut from
the point $q=-1$ is hardly an alternative.

It remains to settle a natural question concerning the case of a homogeneous
elastic material, for which the matrix $\mathbf{Q}$ is constant, hence $%
\mathbf{M}=\exp \left( \mathbf{Q}T\right) ,$ and so $\ln \mathbf{M}$ merely
returns the 'initial' $\mathbf{Q}T,$ which is certainly continuous in $%
\omega .$ 'Technically', the difference with the case of a periodic medium
is that a constant $\mathbf{Q}$ keeps $\mathbf{M}\left( \omega _{0}\right) $
diagonalizable (semisimple) at the degeneracy point $q_{1}\left( \omega
_{0}\right) =q_{2}\left( \omega _{0}\right) =-1$ under discussion\footnote{%
For a constant $\mathbf{Q,}$ this degeneracy of $q_{1,2}=e^{ik_{y}T}$
implies nothing more than an odd number of half-wavelengths within the
interval $\Delta y=T$ - note no relevance to degenerate eigenvalues $k_{y}$
of $\mathbf{Q}$ that do render $\mathbf{Q}$ and hence $\mathbf{M}=\exp
\left( \mathbf{Q}T\right) $ non-semisimple.}. Assuming $\mathbf{M}\left(
\omega _{0}\right) =q_{d}\mathbf{I}$ in Eq. (\ref{13}), its first term turns
to zero and thus a continuous $\ln \mathbf{M}\left( \omega _{0}+\Delta
\omega \right) $ is defined by the second term of (\ref{13}), in which $%
\mathbf{\Delta M}\sim \left( \omega -\omega _{0}\right) $ and $\delta q\sim
\left( \omega -\omega _{0}\right) $ (the latter being due to $B=0$ in (\ref%
{14}) for a semisimple $\mathbf{M}\left( \omega _{0}\right) $ \cite{WaMot2}%
). A transition to (or from) a homogeneous material from (or to) a weakly
(periodically) inhomogeneous one is also evident: given a small parameter $%
\epsilon $ of elastic inhomogeneity, $\mathbf{M}\left( \omega _{0}\right)
-q_{d}\mathbf{I}$ is scaled by $\epsilon $ and $\delta q$ is scaled by $%
\left( \epsilon \Delta \omega \right) ^{1/2},$ hence, by (\ref{13}), the
singularity of $\ln \mathbf{M}\left( \omega \right) $ at $\omega \rightarrow
\omega _{0}$ is proportional to $\left( \epsilon /\Delta \omega \right)
^{1/2}$ and disappears at $\epsilon =0.$

In conclusion, let us outline some exceptional cases that are theoretically
possible due to 'incidental' occurrence of $\mathbf{M}\left( \omega
_{0}\right) $ in a peculiar form. First, a non-semisimple $\mathbf{M}\left(
\omega _{0}\right) $ does not preclude vanishing of the leading-order
coefficient $B$ (\ref{14})$_{2}$ \cite{WaMot2}; if it happens to be zero
then $\left( \delta q\right) ^{2}$ is given by the higher-order terms of the
Taylor series of $D\left( q,\omega \right) $ about $\omega _{0}$, in which
case Eq. (\ref{13}) (where $\mathbf{M}\left( \omega _{0}\right) \neq q_{d}%
\mathbf{I}$) leads to $\ln \mathbf{M}\left( \omega \right) \sim \left(
\omega -\omega _{0}\right) ^{-m/2}$ with an integer $m\geq 2.$ Secondly, a
(periodically) inhomogeneous medium $\mathbf{M}\left( \omega _{0}\right) $
does not rule out a possibility for $\mathbf{M}\left( \omega _{0}\right) $
at a degeneracy point to remain semisimple (such an option is usually
associated with a stopband of zero width). Finally, a semisimple $\mathbf{M}%
\left( \omega _{0}\right) $ may, in principle, also cause diverging $\ln 
\mathbf{M}\left( \omega _{0}+\Delta \omega \right) $ - it is the case when $%
\delta q\sim \left( \omega -\omega _{0}\right) ^{1+\left( m/2\right) }$ with 
$m>0$ due to incidentally vanishing higher-order derivatives $\partial
^{2}D/\partial \omega ^{2},\partial ^{2}D/\partial q\partial \omega $ etc.
in the Taylor series of $D\left( q,\omega \right) $ about $\omega _{0}$,
whence $\ln \mathbf{M}\left( \omega \right) $ for $\omega \rightarrow \omega
_{0}$ diverges owing to the term $\left( \delta q\right) ^{-1}\mathbf{\Delta
M}\sim $ $\left( \omega -\omega _{0}\right) ^{-m/2}$ in Eq. (\ref{13})$.$

\section{Examples of $\mathbf{Q}_{\mathrm{eff}}=i\mathbf{K}$}

\subsection{Bilayered unit cell}

This section is intended to illuminate the preceding general development by
way of its application to simple examples of a scalar acoustic wave in a
periodically repeated sequence of pairs of homogeneous layers. With this
purpose, we first remind the 2$\times $2 setup for an arbitrary 1D-periodic
medium \cite{B,MW} and detail the formulas describing the 'effective' matrix 
$\mathbf{Q}_{\mathrm{eff}}=i\mathbf{K}$ for this framework. Then we further
elaborate $\mathbf{Q}_{\mathrm{eff}}$ for the case of a bilayered unit cell.

\subsubsection{2$\times $2 setup}

Consider a 2$\times $2 unimodular monodromy matrix $\mathbf{M}\left( \omega
\right) \mathbf{.}$ Its eigenvalues%
\begin{equation}
q_{1,2}=\frac{1}{2}\mathrm{trace\,}\mathbf{M}\pm R,\ \mathrm{where\ }R\equiv 
\frac{1}{2}\sqrt{\left( \mathrm{trace\,}\mathbf{M}\right) ^{2}-4}\left( =%
\frac{q_{1}-q_{2}}{2}\right) ,  \label{16}
\end{equation}%
define the Floquet wavenumbers 
\begin{equation}
iK_{1,2}T=\pm iKT=\ln q_{1,2}=\pm i\arccos \left( \frac{1}{2}\mathrm{trace\,}%
\mathbf{M}\right) =\pm 2i\arccos \left( \frac{1}{2}\sqrt{\mathrm{trace\,}%
\mathbf{M}+2}\right) ;  \label{16.1}
\end{equation}%
and the equation 
\begin{equation}
\mathrm{trace\,}\mathbf{M=}-2  \label{17}
\end{equation}%
defines the set of frequencies $\omega =\omega _{0}$ of passband/stopband
crossovers at the BZ edges $KT=\pm \pi $ where $q_{1}\left( \omega
_{0}\right) =q_{2}\left( \omega _{0}\right) \equiv q_{d}=-1,$ see \cite{B,MW}%
.

Introduce the 2$\times $2 'effective' matrix $\mathbf{Q}_{\mathrm{eff}}=i%
\mathbf{K,}$ which is related to $\mathbf{M}$ by the equality $\mathbf{M}%
=\exp \left( i\mathbf{K}T\right) $ and which has eigenvalues (\ref{16.1})
understood under the standard definition of the functions $\ln $ and $%
\arccos ,$ so that $\func{Re}KT\in \left[ -\pi ,\pi \right] .$ Then Eq. (\ref%
{11}) specified for $f\left( \mathbf{M}\right) \equiv \ln \mathbf{M}$ gives 
\begin{equation}
\mathbf{Q}_{\mathrm{eff}}\left( \omega \right) =\frac{iK}{R}\left[ \mathbf{M-%
}\frac{1}{2}\left( \mathrm{trace\,}\mathbf{M}\right) \mathbf{I}\right] .
\label{N4.0}
\end{equation}
The same result may certainly be obtained by equating $\mathbf{M}$ to $\exp
\left( i\mathbf{K}T\right) ,$ which follows from the same (\ref{11})
(re-adjusted to $f\left( \mathbf{K}\right) $) in the form 
\begin{equation}
\exp \left( i\mathbf{K}T\right) =\left( \cos KT\right) \mathbf{I}+\left( i%
\frac{\sin KT}{K}\right) \mathbf{K=}\frac{1}{2}\left( \mathrm{trace\,}%
\mathbf{M}\right) \mathbf{I}+\frac{R}{K}\mathbf{K}  \label{N2}
\end{equation}%
due to using the condition $K_{1,2}=\pm K$ equivalent to fixing the
appropriate definition of matrix logarithm $\ln \mathbf{M}.$

Consider now a vicinity of the BZ edge. Eqs. (\ref{16}), (\ref{16.1}) expand
in small $\Delta \omega =\omega -\omega _{0}$ as%
\begin{equation}
\begin{array}{c}
q_{1,2}\left( \omega \right) \mid _{\omega \approx \omega _{0}}=-1\pm \sqrt{%
B\Delta \omega }+O\left( \Delta \omega \right) ,\ K\left( \omega \right)
T\mid _{\omega \approx \omega _{0}}=\pi +i\sqrt{B\Delta \omega }+O\left(
\Delta \omega \right) , \\ 
R\left( \omega \right) \mid _{\omega \approx \omega _{0}}=\sqrt{B\Delta
\omega +O\left[ \left( \Delta \omega \right) ^{2}\right] },%
\end{array}
\label{N6}
\end{equation}%
where it is denoted 
\begin{equation}
B=-\left( \frac{\mathrm{d}}{\mathrm{d}\omega }\mathrm{trace\,}\mathbf{M}%
\right) _{\omega _{0}},  \label{N7.0}
\end{equation}%
which is non-zero for a non-semisimple $\mathbf{M}\left( \omega _{0}\right) $
(barring the theoretical exceptions mentioned in the end of \S 3.2).
Inserting (\ref{N6}) and (\ref{N7.0}) in (\ref{N4.0}) yields%
\begin{equation}
\mathbf{Q}_{\mathrm{eff}}\left( \omega \right) _{\omega \approx \omega _{0}}=%
\frac{i\pi -\sqrt{B\Delta \omega }+O\left( \Delta \omega \right) }{\sqrt{%
B\Delta \omega +O\left[ \left( \Delta \omega \right) ^{2}\right] }}\left\{ 
\mathbf{A}+\left[ \left( \frac{\mathrm{d}\mathbf{M}}{\mathrm{d}\omega }%
\right) _{\omega _{0}}+\frac{1}{2}B\mathbf{I}\right] \Delta \omega +\mathbf{O%
}\left( \Delta \omega \right) ^{2}\right\} ,  \label{N10}
\end{equation}%
where $\mathbf{A}$ denotes a non-zero nilpotent matrix%
\begin{equation}
\mathbf{A=M}\left( \omega _{0}\right) -q_{d}\mathbf{I=M}\left( \omega
_{0}\right) +\mathbf{I\ \ (A}^{2}=\mathbf{0).\ }  \label{N8.0}
\end{equation}%
Eq. (\ref{N10}) elaborates (\ref{13}) (with due regard for $\mathbf{\Delta M/%
}\delta q\sim \mathbf{O}\left( \delta q\right) $). Note also that Eq. (\ref%
{N6})$_{3}$ for $R,$ defined in (\ref{16})$_{2},$ to leading order reads $%
\delta q=\sqrt{B\Delta \omega }$ which is recognized as the equation (\ref%
{14})$_{1}$. Correspondingly, the definition (\ref{N7.0}) of the coefficient 
$B$ is equivalent to Eq. (\ref{14})$_{2},$ which specializes for the given
case (of 2$\times $2 $\mathbf{M}$ with $q_{d}=-1$ at $\omega _{0}$) as 
\begin{equation}
B=-\left[ \frac{\mathrm{d}}{\mathrm{d}\omega }\det \left( \mathbf{M}-q%
\mathbf{I}\right) \right] _{\omega _{0}}=\mathrm{trace}\left[ \mathbf{A}%
\left( \frac{\mathrm{d}\mathbf{M}}{\mathrm{d}\omega }\right) _{\omega _{0}}%
\right] .  \label{N8}
\end{equation}

Expansion (\ref{N10}) shows that the 'effective' matrix $\mathbf{Q}_{\mathrm{%
eff}}\left( \omega \right) $ has well-behaved eigenvalues $\pm iK\left(
\omega \right) \rightarrow \pm i\pi /T$ at $\omega \rightarrow \omega _{0},$
while its components diverge due to non-zero $\mathbf{A}$ with a common
factor $\sim \left( \omega -\omega _{0}\right) ^{-1/2}$.\ It is also seen
from Eqs. (\ref{N10})-(\ref{N8}) that $\mathbf{A}$ and $B$ for a weakly
inhomogeneous unit cell can in general be scaled by the same small parameter 
$\epsilon $ ($=0$ for a homogeneous limit), and so the singularity of $%
\mathbf{Q}_{\mathrm{eff}}\left( \omega \right) $ at $\omega \rightarrow
\omega _{0}$ is scaled by $\left( \epsilon /\Delta \omega \right) ^{1/2}$ as
argued in \S 3.

\subsubsection{$\mathbf{Q}_{\mathrm{eff}}\ $for a bilayered unit cell}

Let us narrow our analysis to the case of a two-component piecewise constant
unit cell. Specifically, we consider the shear horizontal (SH) wave in a
periodic structure of perfectly bonded pairs of isotropic homogeneous
infinite layers $j=1,2$, each with constant density $\rho _{j},$ shear
modulus $\mu _{j}$ and thickness $d_{j}.$ For the sake of the brevity of
explicit formulas, assume the wave $u\left( y\right) $ propagating along the
axis $Y$ normal to the interfaces ($k_{x}=0$). Hooke's law $\sigma \left(
y\right) =\mu _{j}u^{\prime }\left( y\right) $ and the equation of motion $%
\sigma ^{\prime }\left( y\right) =-\rho _{j}\omega ^{2}u\left( y\right) $
combine into the system (\ref{1}) with the state vector $\mathbf{\eta }%
\left( y\right) =\left( i\omega u,\ \sigma \right) ^{\mathrm{T}}$ and the
piecewise-constant periodic 2$\times $2 system matrix 
\begin{equation}
\mathbf{Q}_{j}=i\omega s_{j}\left( 
\begin{array}{cc}
0 & Z_{j}^{-1} \\ 
Z_{j} & 0%
\end{array}%
\right) ,\ j=1,2,  \label{15.0}
\end{equation}%
which leads to the propagator $\mathbf{M}\left( T,0\right) =\mathrm{e}^{%
\mathbf{Q}_{2}d_{2}}\mathrm{e}^{\mathbf{Q}_{1}d_{1}}\equiv \mathbf{M}\left(
\omega \right) $ through the period $T=d_{1}+d_{2}$ (the monodromy matrix)
in the form 
\begin{equation}
\mathbf{M}\left( \omega \right) =\left( 
\begin{array}{cc}
\cos \psi _{2}\cos \psi _{1}-\frac{Z_{1}}{Z_{2}}\sin \psi _{2}\sin \psi _{1}
& \frac{i}{Z_{1}}\cos \psi _{2}\sin \psi _{1}+\frac{i}{Z_{2}}\sin \psi
_{2}\cos \psi _{1} \\ 
iZ_{1}\cos \psi _{2}\sin \psi _{1}+iZ_{2}\sin \psi _{2}\cos \psi _{1} & \cos
\psi _{2}\cos \psi _{1}-\frac{Z_{2}}{Z_{1}}\sin \psi _{2}\sin \psi _{1}%
\end{array}%
\right) ,  \label{15}
\end{equation}%
where $s_{j}=\sqrt{\rho _{j}/\mu _{j}}$ is the slowness, $Z_{j}=\sqrt{\rho
_{j}\mu _{j}}$ the impedance and $\psi _{j}=\omega s_{j}d_{j}$ the phase
shift over a layer. Passing in (\ref{15}) to an oblique propagation amounts
to merely premultiplying $\psi _{j}$ and $Z_{j}$ by $\sqrt{%
1-s_{x}^{2}/s_{j}^{2}}$ with a fixed $s_{x}=k_{x}/\omega $. Inserting $%
\mathbf{M}$ into the basic relations (\ref{16})-(\ref{17}) provides the
textbook equations implicitly defining the Floquet spectrum $\omega \left(
K\right) $ and its stopband bounds $\omega =\omega _{0}$ at the BG edge for
a bilayered unit cell, e.g.\cite{B}.

The 2$\times $2 'effective' matrix $\mathbf{Q}_{\mathrm{eff}}=i\mathbf{K}$
for a bilayered unit cell follows from (\ref{N4.0}) and (\ref{15}) in the
form%
\begin{equation}
\mathbf{Q}_{\mathrm{eff}}\left( \omega \right) =\frac{iK}{R}%
\begin{pmatrix}
-\frac{1}{2}\left( \frac{Z_{1}}{Z_{2}}-\frac{Z_{2}}{Z_{1}}\right) \sin \psi
_{2}\sin \psi _{1} & \frac{i}{Z_{1}}\cos \psi _{2}\sin \psi _{1}+\frac{i}{%
Z_{2}}\sin \psi _{2}\cos \psi _{1} \\ 
{i}{Z_{1}}\cos \psi _{2}\sin \psi _{1}+{i}{Z_{2}}\sin \psi _{2}\cos \psi _{1}
& \frac{1}{2}\left( \frac{Z_{1}}{Z_{2}}-\frac{Z_{2}}{Z_{1}}\right) \sin \psi
_{2}\sin \psi _{1}%
\end{pmatrix}%
.  \label{N4}
\end{equation}%
It is easy to check that the eigenvalues of this matrix are $\pm iK$, and
that it reduces to (\ref{15.0})$_{1}$ when $Z_{1}=Z_{2},$ $s_{1}=s_{2}.$ As
another consistency test, we note that (\ref{N4}) provides the well-known
low-frequency asymptotics of $\mathbf{Q}_{\mathrm{eff}},$ whose diagonal and
off-diagonal components expand in, respectively, even and odd powers of $%
i\omega $ as follows:%
\begin{equation}
\begin{array}{c}
\mathbf{Q}_{\mathrm{eff}}\left( \omega \right) _{\omega /\omega _{0}\ll
1}=\left\langle \mathbf{Q}\right\rangle +\frac{d_{1}d_{2}}{2T}\left( \mathbf{%
Q}_{2}\mathbf{Q}_{1}-\mathbf{Q}_{1}\mathbf{Q}_{2}\right) +... \\ 
=i\omega \left( 
\begin{array}{cc}
0 & \left\langle \mu ^{-1}\right\rangle \\ 
\left\langle \rho \right\rangle & 0%
\end{array}%
\right) +\frac{1}{2}\left( i\omega \right) ^{2}\kappa T\left( 
\begin{array}{cc}
1 & 0 \\ 
0 & -1%
\end{array}%
\right) +...,%
\end{array}
\label{N5}
\end{equation}%
where%
\begin{equation}
\begin{array}{c}
\mathrm{\ }\left\langle \mathbf{Q}\left( \omega \right) \right\rangle =%
\mathbf{Q}_{1}\frac{d_{1}}{T}+\mathbf{Q}_{2}\frac{d_{2}}{T},\ \left\langle
\mu ^{-1}\right\rangle =\frac{1}{\mu _{1}}\frac{d_{1}}{T}+\frac{1}{\mu _{2}}%
\frac{d_{2}}{T}, \\ 
\ \left\langle \rho \right\rangle =\rho _{1}\frac{d_{1}}{T}+\rho _{2}\frac{%
d_{2}}{T},\ \kappa =\frac{d_{1}d_{2}}{T^{2}}\left( \frac{\rho _{1}}{\mu _{2}}%
-\frac{\rho _{2}}{\mu _{1}}\right) .%
\end{array}
\label{N5.1}
\end{equation}

Additional explicit insight is gained by noticing that $\mathrm{trace\,}%
\mathbf{M}+2$ with $\mathbf{M}$ given by (\ref{15}) can be factored as 
\begin{equation}
\begin{array}{c}
\mathrm{trace\,}\mathbf{M}+2=f_{+}f_{-}, \\ 
f_{\pm }=\frac{1}{\sqrt{Z_{1}Z_{2}}}\left[ \left( Z_{1}+Z_{2}\right) \cos 
\frac{\psi _{1}+\psi _{2}}{2}\pm \left( Z_{1}-Z_{2}\right) \cos \frac{\psi
_{1}-\psi _{2}}{2}\right] ,%
\end{array}
\label{N1}
\end{equation}%
whence Eqs. (\ref{16}), (\ref{16.1}) provide 
\begin{equation}
iK\left( \omega \right) T=\ln \left( \frac{f_{+}f_{-}}{2}-1+R\right)
=2i\arccos \frac{\sqrt{f_{+}f_{-}}}{2},\ R=\sqrt{f_{+}f_{-}\left( \frac{%
f_{+}f_{-}}{4}-1\right) },  \label{N3}
\end{equation}%
and Eq. (\ref{17}) takes the form 
\begin{equation}
f_{+}f_{-}=0,  \label{N1.1}
\end{equation}%
showing that the set $\omega =\omega _{0}$ consists of two families given by
zeros of $f_{\pm }.$ Evidently, this split reveals the
symmetric/antisymmetric decoupling of the problem. As a result, the
expansion (\ref{N10}) of $\mathbf{Q}_{\mathrm{eff}}\left( \omega \right) $
about the points $\omega =\omega _{0},$ when applied to the matrix $\mathbf{Q%
}_{\mathrm{eff}}$ (\ref{N4}) in hand, admits compact formulas for its
leading-order parameters $B$ (\ref{N7.0}) and $\mathbf{A}$ (\ref{N8.0}) as
follows:%
\begin{equation}
\begin{array}{c}
B=-\left( f_{\mp }\frac{\mathrm{d}f_{\pm }}{\mathrm{d}\omega }\right)
_{\omega _{0}}=\mp \frac{1}{\omega _{0}}\left( \frac{Z_{1}}{Z_{2}}-\frac{%
Z_{2}}{Z_{1}}\right) \left( \psi _{1}\sin \psi _{2}+\psi _{2}\sin \psi
_{1}\right) , \\ 
\mathbf{A=\pm }\left( 
\begin{array}{cc}
\cos \psi _{1}+\cos \psi _{2} & \frac{i}{Z_{1}}\sin \psi _{1}-\frac{i}{Z_{2}}%
\sin \psi _{2} \\ 
iZ_{2}\sin \psi _{2}-iZ_{1}\sin \psi _{1} & -\cos \psi _{1}-\cos \psi _{2}%
\end{array}%
\right) ,%
\end{array}
\label{N7}
\end{equation}%
where $\psi _{j}=\omega _{0}s_{j}d_{j}$ are referred to $\omega _{0},$ and
the upper or lower sign corresponds to $f_{+}=0$ or $f_{-}=0$ in (\ref{N1.1}%
), respectively (see \S A3 of Appendix for derivation of (\ref{N7})$_{2}$).
The derivative $\left( \mathrm{d}\mathbf{M/}\mathrm{d}\omega \right)
_{\omega _{0}},$ which also appears in (\ref{N10}), can be obtained due to $%
\mathbf{M}=\mathrm{e}^{\mathbf{Q}_{2}d_{2}}\mathrm{e}^{\mathbf{Q}_{1}d_{1}}$
in the form expressed through the matrices $\mathbf{Q}_{j}$ (\ref{15.0}) and 
$\mathbf{A}$ as%
\begin{equation}
\left( \frac{\mathrm{d}\mathbf{M}}{\mathrm{d}\omega }\right) _{\omega _{0}}=%
\frac{1}{\omega _{0}}\left( d_{2}\mathbf{Q}_{2}\mathbf{M}+d_{1}\mathbf{MQ}%
_{1}\right) _{\omega _{0}}=\frac{T}{\omega _{0}}\left[ \frac{d_{2}}{T}%
\mathbf{Q}_{2}\left( \omega _{0}\right) \mathbf{A}+\frac{d_{1}}{T}\mathbf{AQ}%
_{1}\left( \omega _{0}\right) -\left\langle \mathbf{Q}\left( \omega
_{0}\right) \right\rangle \right] .  \label{N8.1}
\end{equation}%
Its plugging in (\ref{N8})$_{2}$ and taking note of $\mathbf{A}^{2}=\mathbf{0%
}$ yields another definition of the coefficient $B$,%
\begin{equation}
B=-\frac{T}{\omega _{0}}\mathrm{trace}\left[ \mathbf{A}\left\langle \mathbf{Q%
}\left( \omega _{0}\right) \right\rangle \right] \mathrm{\,},  \label{N9}
\end{equation}%
which for the given case of a bilayered unit cell is equivalent to (\ref%
{N7.0}) and (\ref{N8}). It is easy to verify that (\ref{N9}) with (\ref{N5.1}%
) and (\ref{N7})$_{2}$ leads to (\ref{N7})$_{1}.$

The following analysis for highly contrasting layers and for layers in
spring-mass-spring contact makes an extensive use of the factorization (\ref%
{N1}) and the consequent formulas.

\subsubsection{High-contrast case}

It is instructive to specialize the above considerations to the case of high
contrast between the material properties of two layers composing the unit
cell. Suppose that, e.g., the second layer is much softer than the first
one: 
\begin{equation}
\mu _{2}/\mu _{1}\equiv \varepsilon ^{2}\ \left( =>s_{2}\sim \varepsilon
^{-1},\ Z_{2}\sim \varepsilon \right) ,\ \mathrm{where\ }0<\varepsilon \ll 1.
\label{18}
\end{equation}%
The main interest of the high-contrast case is that the first stopband at
the BZ edge occurs in the low-frequency range, which is scaled by $%
\varepsilon $ and implies $\psi _{1}=O\left( \varepsilon \right) ,\ \psi
_{2}=O\left( 1\right) .$ In this range, the propagator (\ref{15}) is
approximated to leading order in $\varepsilon $ as%
\begin{equation}
\mathbf{M}\left( \omega \right) _{\psi _{1}=O\left( \varepsilon \right)
}=\left( 
\begin{array}{cc}
\cos \psi _{2}-\beta \sin \psi _{2} & \frac{i}{Z_{2}}\sin \psi _{2} \\ 
iZ_{2}\left( \sin \psi _{2}+\beta \cos \psi _{2}\right) & \cos \psi _{2}%
\end{array}%
\right) \ ,  \label{19}
\end{equation}%
where 
\begin{equation}
\beta \left( \omega \right) \equiv \frac{Z_{1}\psi _{1}}{Z_{2}}\left(
=\omega \frac{\rho _{1}d_{1}}{\sqrt{\rho _{2}\mu _{2}}}=\frac{\rho _{1}d_{1}%
}{\rho _{2}d_{2}}\psi _{2}\right) ;  \label{21}
\end{equation}%
and Eq. (\ref{17}) with $\mathbf{M}$ (\ref{19}) defines the stopband bounds $%
\omega =\omega _{0}$ by 
\begin{equation}
\cos \psi _{2}-\frac{\beta }{2}\sin \psi _{2}=-1\ \Leftrightarrow \cos \frac{%
\psi _{2}}{2}\left( \cos \frac{\psi _{2}}{2}-\frac{\beta }{2}\sin \frac{\psi
_{2}}{2}\right) =0.\   \label{20}
\end{equation}%
The latter, factorized, form is Eq. (\ref{N1.1}) with approximate $f_{\pm }$
(\ref{N1})$_{2}.$ So the first stopband is bounded by the least roots of $%
f_{+}=0$ and $f_{-}=0$ which, to leading order in $\varepsilon ,$ are the
first zeros of the cofactors of (\ref{20})$_{2}.$ The upper bound
corresponding to $f_{+}=0$ is close to the first thickness resonance $\psi
_{2}\left( =\omega s_{2}d_{2}\right) =\pi $ of the soft layer. Denote the
lower bound corresponding to $f_{-}=0$ by $\Omega ~\left( =\min \omega
_{0}\right) .$ It is approximated by the least root of equation 
\begin{equation}
\tan \left( \psi _{2}/2\right) =2/\beta ,  \label{20.1}
\end{equation}%
which involves coupling of the layers. Note in passing resemblance and
dissimilarity between this simple model (see also \S 4.2) and the textbook
case of a high-contrast diatomic lattice \cite{B}.

With a view to highlight the low-frequency behaviour of $\mathbf{Q}_{\mathrm{%
eff}}\left( \omega \right) $, let us focus our attention on $\omega $
ranging from $\omega \approx \Omega $ and going down the first Floquet
branch to $\omega =0.$ Substituting (\ref{19}) in (\ref{N4}) yields%
\begin{equation}
\mathbf{Q}_{\mathrm{eff}}\left( \omega \right) =\frac{iK}{R}\left( 
\begin{array}{cc}
-\frac{\beta }{2}\sin \psi _{2} & \frac{i}{Z_{2}}\sin \psi _{2} \\ 
iZ_{2}\left( \sin \psi _{2}+\beta \cos \psi _{2}\right) & \frac{\beta }{2}%
\sin \psi _{2}%
\end{array}%
\right) ,  \label{26}
\end{equation}%
where by (\ref{N3}) and (\ref{20}) 
\begin{equation}
\begin{array}{c}
iK\left( \omega \right) T=\ln \left( \cos \psi _{2}-\frac{\beta }{2}\sin
\psi _{2}+R\right) =2i\arccos \left[ \cos \frac{\psi _{2}}{2}\left( \cos 
\frac{\psi _{2}}{2}-\frac{\beta }{2}\sin \frac{\psi _{2}}{2}\right) \right]
^{1/2}, \\ 
R\left( \omega \right) =\left[ 2\sin \psi _{2}\left( \frac{\beta }{2}\cos 
\frac{\psi _{2}}{2}+\sin \frac{\psi _{2}}{2}\right) \left( \frac{\beta }{2}%
\sin \frac{\psi _{2}}{2}-\cos \frac{\psi _{2}}{2}\right) \right] ^{1/2}.%
\end{array}
\label{24}
\end{equation}%
The singular term for $\mathbf{Q}_{\mathrm{eff}}\left( \omega \right) $ (\ref%
{26}) as $\omega $ tends to the first stopband bound $\Omega $ is $\mathbf{Q}%
_{\mathrm{eff}}\left( \omega \right) \varpropto \frac{i\pi }{\sqrt{B\Delta
\omega }}\mathbf{A}$ (see (\ref{N10})) with $\sqrt{\Delta \omega }=i\sqrt{%
\Omega -\omega }$ and 
\begin{equation}
B=\frac{1}{\Omega }\left[ \beta \left( \psi _{2}+\sin \psi _{2}\right) %
\right] _{\omega =\Omega },\ \mathbf{A}=\frac{2}{1+\left( 2/\beta \right)
^{2}}\left( 
\begin{array}{cc}
-1 & \frac{2i}{Z_{2}\beta } \\ 
iZ_{2}\beta /2 & 1%
\end{array}%
\right) _{\omega =\Omega }\mathbf{.}  \label{22}
\end{equation}%
Eq. (\ref{22}) follows from (\ref{N7}) which is taken with the lower sign
(since $\Omega $ is defined by $f_{-}=0$) and confined to leading order in $%
\varepsilon $ (in accordance with the accuracy of (\ref{19}) and hence of (%
\ref{26})). The asymptotics of the same $\mathbf{Q}_{\mathrm{eff}}\left(
\omega \right) $ (\ref{26}) near the origin point $\omega =0$ is given by
Eq. (\ref{N5}) with%
\begin{equation}
\left\langle \mu ^{-1}\right\rangle =\frac{1}{\mu _{2}}\frac{d_{2}}{T},\
\kappa =\frac{d_{1}d_{2}}{T^{2}}\frac{\rho _{1}}{\mu _{2}},  \label{23}
\end{equation}%
which also implies taking leading order in the high-contrast parameter $%
\varepsilon .$ Note that $B$ provided in (\ref{22})$_{1}$ satisfies Eq. (\ref%
{N9}) with $\left\langle \mathbf{Q}\right\rangle $ given by (\ref{23}).

\subsection{Layers in spring-mass-spring contact}

As another example, consider propagation of the SH wave through a structure
of identical layers of thickness $T$ in spring-mass-spring contact. Denote
the rigidity of each of two springs by $\gamma $ and the mass by $m.$ Note
that the physical dimension of $m$ is voluminal density times length. The
monodromy matrix $\mathbf{M\equiv M}\left( T,0\right) $ for the state vector 
$\mathbf{\eta }=\left( i\omega u,\ \sigma \right) ^{\mathrm{T}}$ is $\mathbf{%
M}=\mathbf{M}_{\mathrm{int}}\mathbf{M}_{l},$ where $\mathbf{M}_{l}=\exp
\left( \mathbf{Q}T\right) $ is the propagator across the layer with $\mathbf{%
Q}$ given by (\ref{15.0}) (no subscripts $j=1,2$), and $\mathbf{M}_{\mathrm{%
int}}$ is the propagator across the spring-mass-spring interface:%
\begin{equation}
\mathbf{M}_{\mathrm{int}}=\left( 
\begin{array}{cc}
1-\frac{\omega ^{2}m}{\gamma } & \frac{2i\omega }{\gamma }\left( 1-\frac{%
\omega ^{2}m}{2\gamma }\right) \\ 
i\omega m & 1-\frac{\omega ^{2}m}{\gamma }%
\end{array}%
\right) =\left( 
\begin{array}{cc}
1-\frac{2\omega ^{2}}{\Omega _{r}^{2}} & \frac{2i\omega }{\gamma }\left( 1-%
\frac{\omega ^{2}}{\Omega _{r}^{2}}\right) \\ 
\frac{2i\omega \gamma }{\Omega _{r}^{2}} & 1-\frac{2\omega ^{2}}{\Omega
_{r}^{2}}%
\end{array}%
\right) ,  \label{27}
\end{equation}%
where $\Omega _{r}=\sqrt{2\gamma /m}$ is the resonant frequency of this
joint. Thus%
\begin{equation}
\mathbf{M}\left( \omega \right) =\left( 
\begin{array}{cc}
\left( 1-\frac{2\omega ^{2}}{\Omega _{r}^{2}}\right) \cos \psi -\frac{%
2\omega Z}{\gamma }\left( 1-\frac{\omega ^{2}}{\Omega _{r}^{2}}\right) \sin
\psi & \frac{i}{Z}\left( 1-\frac{2\omega ^{2}}{\Omega _{r}^{2}}\right) \sin
\psi +\frac{2i\omega }{\gamma }\left( 1-\frac{\omega ^{2}}{\Omega _{r}^{2}}%
\right) \cos \psi \\ 
iZ\left( 1-\frac{2\omega ^{2}}{\Omega _{r}^{2}}\right) \sin \psi +\frac{%
2i\omega \gamma }{\Omega _{r}^{2}}\cos \psi & \left( 1-\frac{2\omega ^{2}}{%
\Omega _{r}^{2}}\right) \cos \psi -\frac{2\omega \gamma }{Z\Omega _{r}^{2}}%
\sin \psi%
\end{array}%
\right) .  \label{28}
\end{equation}%
A factorized form (\ref{N1.1}) of the equation (\ref{17}) defining the
stopbands at the edge of the BZ holds with%
\begin{equation}
\begin{array}{c}
f_{+}=4\left[ \left( 1-\psi ^{2}\frac{m}{\rho T}\frac{\mu }{2\gamma T}%
\right) \cos \frac{\psi }{2}-\frac{m}{2\rho T}\psi \sin \frac{\psi }{2}%
\right] , \\ 
f_{-}=\cos \frac{\psi }{2}-\frac{\mu }{\gamma T}\psi \sin \frac{\psi }{2},%
\end{array}
\label{29}
\end{equation}%
where $\omega ^{2}/\Omega _{r}^{2}=\psi ^{2}m\mu /2\rho \gamma T^{2}$ is
used to write $f_{\pm }$ as functions of the phase shift $\psi =\omega T%
\sqrt{\rho /\mu }.$ It is seen that $f_{+}\left( \psi \right) $ depends on
both spring and mass parameters $\gamma T/\mu $ and $m/\rho T,$ while $%
f_{-}\left( \psi \right) $ depends on $\mu /\gamma T$ only.

Let us again specialize our consideration to the high-contrast case of a
similar 'stiff/soft' nature, now by assuming a relatively small rigidity 
\begin{equation}
\gamma T/\mu \ll 1  \label{30}
\end{equation}%
of the springs supporting the mass. Like before, we are interested in the
first stopband at the BZ edge. Given (\ref{30}), the least roots $\psi _{\pm
}$ of $f_{\pm }=0$ and the corresponding stopband bounds $\Omega _{\pm }$ to
leading order are 
\begin{equation}
\begin{array}{c}
\psi _{+}=\min \left( \sqrt{\frac{2\gamma T}{\mu }\frac{\rho T}{m}},~\pi
\right) \ \Rightarrow \ \Omega _{+}=\min \left( \Omega _{r},\ \Omega
_{l}\right) =\min \left( \sqrt{\frac{2\gamma }{m}},\ \frac{\pi }{T}\sqrt{%
\frac{\mu }{\rho }}\right) ; \\ 
\psi _{-}=\sqrt{\frac{2\gamma T}{\mu }}\ \Rightarrow \ \Omega _{-}=\sqrt{%
\frac{2\gamma }{\rho T}},%
\end{array}%
\   \label{31}
\end{equation}%
where $\Omega _{l}$ is the frequency of the thickness resonance of the
layer. The question is which of $\Omega _{+}$ and $\Omega _{-}$ is the lower
frequency bound. Since $\Omega _{-}^{2}/\Omega _{r}^{2}=m/\rho T,$ it is
evident that a heavy mass $m\gg \rho T$ ensures $\Omega _{+}=\Omega
_{r}<\Omega _{-};$ a 'medium heavy' mass $m\sim \rho T$ implies commensurate 
$\Omega _{+}=\Omega _{r}\sim \Omega _{-},$ and a light mass $m\ll \rho T$
ensures $\Omega _{-}<\Omega _{+}.$ For the two former cases, the whole first
stopband is confined to the low-frequency range in the sense that both its
bounds provide a small phase $\psi \ll 1.$ In the latter case of a light
mass, decreasing the small parameter $m/\rho T$ keeps the lower bound at $%
\psi _{-}\ll 1$ and lifts the upper bound up until the phase $\psi _{+}$
reaches $\pi ,$ i.e. $\Omega _{r}$ reaches $\Omega _{l}$.

Consider the range $\psi \ll 1$ containing one or both bounds ($\Omega _{-}$
or $\Omega _{-}$ and $\Omega _{+}=\Omega _{r}$, respectively) of the first
stopband at the BZ edge. Expanding (\ref{28}) to leading order in small $%
\psi ,$ bearing in mind (\ref{30}), and using the notations (\ref{31}) of $%
\Omega _{-}$ and $\Omega _{r}$ yields%
\begin{equation}
\mathbf{M}\left( \omega \right) =\left( 
\begin{array}{cc}
1-\frac{2\omega ^{2}}{\Omega _{r}^{2}}-4\frac{\omega ^{2}}{\Omega _{-}^{2}}%
\left( 1-\frac{\omega ^{2}}{\Omega _{r}^{2}}\right) & \frac{2i\omega }{%
\gamma }\left( 1-\frac{\omega ^{2}}{\Omega _{r}^{2}}\right) \\ 
\frac{2i\omega \gamma }{\Omega _{-}^{2}}\left( 1-\frac{2\omega ^{2}}{\Omega
_{r}^{2}}+\frac{\Omega _{-}^{2}}{\Omega _{r}^{2}}\right) & 1-\frac{2\omega
^{2}}{\Omega _{r}^{2}}%
\end{array}%
\right) ,  \label{32}
\end{equation}%
which observes $\det \mathbf{M}=1$. Inserting (\ref{32}) in (\ref{N4.0})
gives 
\begin{equation}
\mathbf{Q}_{\mathrm{eff}}\left( \omega \right) =\frac{iK}{R}\left( 
\begin{array}{cc}
-2\frac{\omega ^{2}}{\Omega _{-}^{2}}\left( 1-\frac{\omega ^{2}}{\Omega
_{r}^{2}}\right) & \frac{2i\omega }{\gamma }\left( 1-\frac{\omega ^{2}}{%
\Omega _{r}^{2}}\right) \\ 
\frac{2i\omega \gamma }{\Omega _{-}^{2}}\left( 1-\frac{2\omega ^{2}}{\Omega
_{r}^{2}}+\frac{\Omega _{-}^{2}}{\Omega _{r}^{2}}\right) & 2\frac{\omega ^{2}%
}{\Omega _{-}^{2}}\left( 1-\frac{\omega ^{2}}{\Omega _{r}^{2}}\right)%
\end{array}%
\right) ,  \label{35}
\end{equation}%
in which 
\begin{equation}
\begin{array}{c}
iK\left( \omega \right) T=\ln \left( 2\alpha -1+R\right) =2i\arccos \sqrt{%
\alpha },\ R\left( \omega \right) =2i\sqrt{\alpha \left( 1-\alpha \right) }
\\ 
\mathrm{with}\ \alpha =\left( 1-\omega ^{2}/\Omega _{r}^{2}\right) \left(
1-\omega ^{2}/\Omega _{-}^{2}\right) .%
\end{array}
\label{36}
\end{equation}
The latter follows from a similar expansion of $f_{\pm }\left( \psi \right) $
(\ref{29}) at $\psi \ll 1\ $and $\gamma T/\mu \ll 1,$ approximating the
l.h.s. of Eq. (\ref{N1.1}) as $f_{+}f_{-}=4\alpha $ and plugging it into (%
\ref{N3}).

According to (\ref{35}) and (\ref{36}), the matrix $\mathbf{Q}_{\mathrm{eff}%
}\left( \omega \right) ,$ as expected, experiences the square-root
singularity at the BZ edge; however, it does so in a different way when $%
\omega $ approaches either $\Omega _{-}$ (light mass) or $\Omega _{r}$ (if $%
\Omega _{r}$ fulfils $\psi \ll 1;$ heavy mass). By (\ref{32}) and (\ref{35}%
), all components of the matrix $\mathbf{A=M}\left( \Omega _{-}\right) -%
\mathbf{I}$ are non-zero and hence all components of $\mathbf{Q}_{\mathrm{eff%
}}\left( \omega \right) $ diverge when $\omega \rightarrow \Omega _{-}.$
This is a typical option for a singularity of $\mathbf{Q}_{\mathrm{eff}%
}\left( \omega \right) .$ On the other hand, $\mathbf{A=M}\left( \Omega
_{r}\right) -\mathbf{I}$ has only left off-diagonal component being
non-zero, and hence only this component of $\mathbf{Q}_{\mathrm{eff}}\left(
\omega \right) $ diverges when $\omega \rightarrow \Omega _{r}$ while the
others tend to zero. This is rather an unusual option, which is due to the
approximations underlying a simple form (\ref{32}) and (\ref{35}) of $%
\mathbf{M}$ and $\mathbf{Q}_{\mathrm{eff}}.$ The transition between the two
above options occurs at $\Omega _{-}=\Omega _{r}$ (i.e. $m=\rho T$)$,$ in
which case $\omega _{0}=\Omega _{-}=\Omega _{r}$ implies the stopband of
zero width that yields a semisimple $\mathbf{M}\left( \omega _{0}\right) =-%
\mathbf{I}$ so that $\mathbf{Q}_{\mathrm{eff}}\left( \omega \right) $ is
well-behaved at $\omega \rightarrow \omega _{0}$ (it is one of the
extraordinary possibilities mentioned in the end of \S 3.2). For either of
these cases, the low-frequency asymptotics of $\mathbf{Q}_{\mathrm{eff}%
}\left( \omega \right) $ (\ref{35}) is given by (\ref{N5}) with the
effective properties taken to leading order in the soft-spring parameter (%
\ref{30}), i.e., with%
\begin{equation}
\left\langle \mu ^{-1}\right\rangle =\frac{2}{\gamma T},\ \ \left\langle
\rho \right\rangle =\rho +\frac{m}{T},\ \kappa =\frac{2\rho }{\gamma T},
\label{37}
\end{equation}%
where $\gamma _{s}=\gamma /2$ is the rigidity of two identical springs in
series (cf. (\ref{23})).

%%%%%%%%%%%%%%%%%%%%%%%%%%%%%%%%%%%%%%%%%%%%%%%%%% Figure
\begin{figure*}[htbp]
\centering
\includegraphics[width=5.0in , height=2.5in 					]{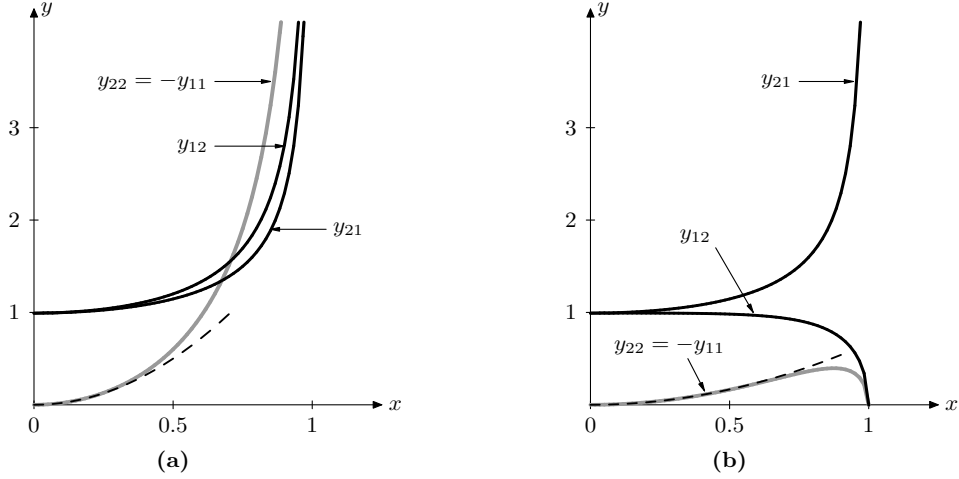} 
	\caption{Frequency dependence of components of the 'effective' matrix $%
\mathbf{Q}_{\mathrm{eff}}$ (\ref{35}) in the first passband, which is (a) $%
\omega \in \left[ 0,\Omega _{-}\right] $ with $\Omega _{-}^{2}/\Omega
_{r}^{2}=1/3$ and (b) $\omega \in \left[ 0,\Omega _{r}\right] $ with $\Omega
_{-}^{2}/\Omega _{r}^{2}=3.$ Black curves are the off-diagonal components $%
ij=12,~21$ normalized by their statically averaged values; grey curves are
diagonal components, whose leading low-frequency evaluation ($\sim \omega
^{2}$) is shown by dashed line. The curves definition is specified in the
text.	} 
\end{figure*}
%%%%%%%%%%%%%%%%%%%%%%%%%%%%%%%%%%%%%%%%%%%%%%%%%% 

The two types of singular behaviour of the 'effective' matrix $\mathbf{Q%
}_{\mathrm{eff}}\left( \omega \right) $ defined by (\ref{35}), (\ref{36})
are illustrated in Fig. 1. It displays the off-diagonal components
normalized by their statically-averaged values $\left\langle \mathbf{Q}%
\right\rangle _{ij}$ ($\sim \omega $, see (\ref{N5}) with (\ref{37})$_{1,2}$%
) and compares the diagonal components to their leading low-frequency term ($%
\sim \omega ^{2},$ see (\ref{N5}) with (\ref{37})$_{3}$). Specifically, the
plotted curves are defined as $y_{ij}\left( x\right) =\frac{1}{\left\langle 
\mathbf{Q}\right\rangle _{ij}}\left( \mathbf{Q}_{\mathrm{eff}}\right) _{ij}$
($ij=12,21$) and $y_{ii}\left( x\right) =T\left( \mathbf{Q}_{\mathrm{eff}%
}\right) _{ii}$ ($y_{22}=-y_{11}$) with $x=\omega /\Omega _{-}$ when $\Omega
_{-}^{2}=\left( 1/3\right) \Omega _{r}^{2}$ (Fig. 1a) and with $x=\omega
/\Omega _{r}$ when $\Omega _{-}^{2}=3\Omega _{r}^{2}$ (Fig. 1b), where $%
\Omega _{-}^{2}/\Omega _{r}^{2}=m/\rho T$ $\left( \gg \gamma T/2\mu \right)
. $

Note in conclusion that passing to the case of an oblique propagation ($%
k_{x}=\omega s_{x}\neq 0,$ see note to (\ref{15})) implies replacing the
entries of layer density $\rho $ by $\rho -s_{x}^{2}\mu $. Moreover, this
case enables further 'ramification' of the spring-mass-spring model by means
of recasting the point mass $m$ as an 'elastic' mass $m\left(
1-c_{T}^{2}s_{x}^{2}\right) $ with its own shear velocity $c_{T}$ (then $%
\Omega _{r}^{2}$ becomes $\Omega _{r}^{2}=2\gamma /m\left(
1-c_{T}^{2}s_{x}^{2}\right) $). It is also noted that the case of layers in
'pure spring' contact (i.e. without a mass) is described by the above
formulas taken with $m=0$ ($\Omega _{r}^{2}\rightarrow \infty $) and with $%
\gamma _{s}=\gamma /2$ as the rigidity of the spring joint, or else by the
formulas of \S 4.1 taken in the limit $d_{2}\rightarrow 0$, $\mu
_{2}\rightarrow 0$ while keeping $\gamma _{s}=\mu _{2}/d_{2}$ finite.

\section{Summary}

Components of the matrix logarithm $\ln \mathbf{M},$ where $\mathbf{M=M}%
\left( T,0\right) $ is a unimodular propagator matrix relating the acoustic
wave field with a frequency $\omega $ at one and the other ends of a period $%
T$ of 1D-periodic anisotropic medium, have been shown to diverge when the
frequency $\omega $ tends to the values $\omega _{0}$ of passband/stopband
crossovers occurring at the edge of the first Brillouin zone (BZ). Explicit
analytical examples of the 'effective' matrix $\mathbf{Q}_{\mathrm{eff}%
}\left( \omega \right) \left( \equiv i\mathbf{K}\left( \omega \right)
\right) =\frac{1}{T}\ln \mathbf{M}\left( \omega \right) $ and of its
diverging asymptotics near the BZ edges were provided for the simple case of
a scalar waves in a two-component periodic structure of several types,
including its high-contrast model when the least of $\omega _{0}$ may be
made arbitrarily small.

Whereas the components of matrix $\mathbf{Q}_{\mathrm{eff}}$ diverge at $%
\omega \rightarrow \omega _{0}$, it is understood that $\mathbf{Q}_{\mathrm{%
eff}}$ for any $\omega \neq \omega _{0}$ yields a continuous $\mathbf{M}%
=\exp \left( \mathbf{Q}_{\mathrm{eff}}T\right) $ and has a continuous
eigenspectrum which is in one-to-one correspondence with that of $\mathbf{M.}
$ Thus, invoking a diverging $\mathbf{Q}_{\mathrm{eff}}$ for formulating a
time-harmonic wave propagation through a finite or infinite number of
periods cannot create any difficulty, because this phenomenon can be fully
described via $\mathbf{M}$ and its eigenspectrum$.$ At the same time,
divergence of components of $\mathbf{Q}_{\mathrm{eff}}$ calls for careful
interpretation if the governing system (\ref{1}) is taken with $\mathbf{Q}_{%
\mathrm{eff}}$ in place of the actual matrix of coefficients $\mathbf{Q}%
\left( y\right) $ and is then viewed in the same sense as the 'true' system (%
\ref{1}), i.e., as incorporating the equation of motion and the constitutive
law, but now with the constant coefficients $\mathbf{Q}_{\mathrm{eff}}\left(
\omega \right) $ of the fictitious homogenized medium.

Explicit results of this paper can readily be adjusted to other physical
problems whose mathematical formulation admits reduction to Eq. (\ref{1}),
see e.g. \cite{ZBG}. Further development is underway to analyze the Floquet
dispersion near an arbitrary passband/stopband crossover occurring anywhere
in the BZ of 1D-periodic structure. Another interest lies in the potential
extension of the analytical means of the paper to more complicated cases,
like in \cite{ACG}, whose exact mathematical statement does not reduce to (%
\ref{1}).\medskip

\noindent \textbf{Acknowledgement.} AAK and ANN wish to express thanks to
the Laboratoire de M\'{e}canique Physique (LMP) of the Universit\'{e}
Bordeaux 1 for the hospitality. The visit of AAK to the LMP has been
supported by the grant ANR-08-BLAN-0101-01 from the Agence Nationale de la
Recherche. The authors are grateful to O. Poncelet for numerous stimulating
discussions.

%\pagebreak

%\pagebreak
\bigskip

\noindent {\large APPENDIX}\textbf{\medskip }

\noindent \textbf{A1}. \textbf{On the divergence of the logarithm of} 
\textbf{a }$n\times n$ \textbf{matrix} $\mathbf{M}(\omega )\smallskip $

Let $\mathbf{M}(\omega )$ be a $n\times n$ non-singular ($\det \mathbf{M}%
\neq 0$) matrix, continuous in $\omega ,$ with eigenvalues $q_{j}(\omega ).$
Denote $\mathbf{M}(\omega _{0})=\mathbf{M}_{0},$ $q_{j}(\omega
_{0})=q_{j}^{0}$ and suppose that $q_{1}^{0}=q_{2}^{0}\equiv q_{d}$ while
all other $q_{j}^{0}$ ($j\neq 1,2$) are distinct. Consider a small
neighbourhood of $\omega _{0},$ where 
\begin{equation}
\mathbf{M}(\omega )=\mathbf{M}_{0}+\mathbf{o}(1),\ q_{j}(\omega
)=q_{j}^{0}+o(1),  \label{A1}
\end{equation}%
and all $q_{j}(\omega )$ ($j=1,2,...,n$) are distinct. Assume that the
matrix $\mathbf{M}_{0}$ with a degenerate eigenvalue $q_{d}$ is
non-semisimple, i.e., that the Jordan form $\mathbf{J}_{0}$ of $\mathbf{M}%
_{0}$ is 
\begin{equation}
\mathbf{J}_{0}=\mathbf{P}\oplus \mathbf{S\ }\mathrm{with}\mathbf{\ P}=q_{d}%
\mathbf{I}_{2}+\mathbf{R,\ R}=\left( 
\begin{array}{cc}
0 & 1 \\ 
0 & 0%
\end{array}%
\right) ,\ \mathbf{S}=\mathrm{diag}\left( q_{3}^{0},...,q_{n}^{0}\right) ,
\label{A2}
\end{equation}%
where $\mathbf{I}_{m}$ denotes the $m\times m$ identity matrix. Thus the
spectral decomposition of $\mathbf{M}(\omega )$ is%
\begin{equation}
\begin{array}{c}
\mathbf{M}(\omega )=\mathbf{C}\left( \omega \right) \mathrm{diag}\left(
q_{1}\left( \omega \right) ,...,q_{n}\left( \omega \right) \right) \mathbf{C}%
^{-1}\left( \omega \right) \ \mathrm{for\ }\omega \neq \omega _{0}, \\ 
\mathbf{M}_{0}=\mathbf{C}_{0}\mathbf{J}_{0}\mathbf{C}_{0}^{-1}\ \mathrm{for\ 
}\omega =\omega _{0},%
\end{array}
\label{A3}
\end{equation}%
where $\mathbf{C}\left( \omega \right) $ and $\mathbf{C}_{0}$ are matrices
whose columns are linear independent eigenvectors of, respectively, $\mathbf{%
M}(\omega )$ and $\mathbf{M}_{0}$ (note that $\mathbf{C}_{0},$ which
includes a generalized eigenvector of $\mathbf{M}_{0},$ is certainly not $%
\mathbf{C}\left( \omega _{0}\right) ,$ which is singular).

Introduce a logarithm of $\mathbf{M}\left( \omega \right) $ with $\omega
\neq \omega _{0},$ 
\begin{equation}
\begin{array}{c}
\ln \mathbf{M}\left( \omega \right) =\mathbf{C}\left( \omega \right) \mathrm{%
diag}\left( \ln q_{1}\left( \omega \right) ,...,\ln q_{n}\left( \omega
\right) \right) \mathbf{C}^{-1}\left( \omega \right) , \\ 
\mathrm{where}\ \ln q_{j}=\ln |q_{j}|+i(\arg q_{j}+2\pi k_{j}),\ k_{j}\in 
\mathbb{Z}.%
\end{array}
\label{A4}
\end{equation}%
This is a general definition in the sense that, while observing indeed the
equality $\exp \left[ \ln \mathbf{M}\left( \omega \right) \right] =\mathbf{M}%
\left( \omega \right) ,$ it permits taking each $\ln q_{j}$ in (\ref{A4}$%
_{2} $) on any $k_{j}$-th Riemann sheet. Let us further suppose that 
\begin{equation}
\ln q_{1}^{0}-\ln q_{2}^{0}=2\pi i,  \label{A5}
\end{equation}%
which implies either that $q_{1}\left( \omega \right) $ and $q_{2}\left(
\omega \right) $ tending to $q_{d}$ as $\omega \rightarrow \omega _{0}$ are
defined on adjacent Riemann sheets ($k_{1}-k_{2}=1$) and $q_{d}$ is away
from the cut, or, alternatively, that $q_{1}\left( \omega \right) $ and $%
q_{2}\left( \omega \right) $ are taken on the same Riemann sheet ($%
k_{1}=k_{2}$ in (\ref{A4}$_{2}$)) with the cut such that $q_{1}^{0}$ and $%
q_{2}^{0}$ are located on its opposite edges. The latter option with $%
k_{1,2}=0$ is directly related to the physical context discussed in this
paper.

Our purpose is to show that, under the aforementioned assumptions, the
asymptotics of $\ln \mathbf{M}\left( \omega \right) $ at $\omega \rightarrow
\omega _{0}$ is%
\begin{equation}
\begin{array}{c}
\ln \mathbf{M}\left( \omega \right) =\frac{2\pi i}{q_{1}\left( \omega
\right) -q_{2}\left( \omega \right) }\mathbf{A}+\mathbf{o}\left( \frac{1}{%
q_{1}\left( \omega \right) -q_{2}\left( \omega \right) }\right) \ \mathrm{%
with\ } \\ 
\mathbf{A}=\mathbf{C}_{0}\left[ \mathbf{J}_{0}-\mathrm{diag}\left(
q_{d},q_{d},q_{3}^{0},...,q_{n}^{0}\right) \right] \mathbf{C}_{0}^{-1}=%
\mathbf{C}_{0}(\mathbf{R}\oplus \mathbf{0}_{n-2})\mathbf{C}_{0}^{-1}\neq 
\mathbf{0}_{n},%
\end{array}
\label{A6}
\end{equation}%
where $\mathbf{0}_{m}$ is $m\times m$ zero matrix and the other entries have
been defined above.

The derivation of (\ref{A6}) is based on the Lagrange-Sylvester formula \cite%
{HJ} with due regard for (\ref{A1}), (\ref{A3}$_{2}$) and (\ref{A5}). Along
these lines, we manipulate $\ln \mathbf{M}\left( \omega \right) $ as follows
(omitting for brevity the argument $\omega $ of $\mathbf{M}\left( \omega
\right) $ and $q_{j}\left( \omega \right) $):%
\begin{equation}
\begin{array}{c}
\ln \mathbf{M}=\sum_{k=1}^{n}\left( \prod_{j\neq k}\frac{\mathbf{M}-q_{j}%
\mathbf{I}_{n}}{q_{k}-q_{j}}\right) \ln q_{k} \\ 
=\sum_{k=1}^{n}\left( \prod_{j\neq k}\frac{\mathbf{M}_{0}-q_{j}^{0}\mathbf{I}%
_{n}}{q_{k}-q_{j}}\right) \ln q_{k}^{0}+\mathbf{o}\left( \frac{1}{q_{1}-q_{2}%
}\right) \\ 
=\sum_{k=1}^{2}\left( \prod_{j\neq k}\frac{\mathbf{M}_{0}-q_{j}^{0}\mathbf{I}%
_{n}}{q_{k}-q_{j}}\right) \ln q_{k}^{0}+\mathbf{o}\left( \frac{1}{q_{1}-q_{2}%
}\right) \\ 
=\left( \prod_{j\geq 3}\frac{\mathbf{M}_{0}-q_{j}^{0}\mathbf{I}_{n}}{%
q_{d}-q_{j}^{0}}\right) \frac{\mathbf{M}_{0}-q_{d}\mathbf{I}_{n}}{q_{1}-q_{2}%
}(\ln q_{1}^{0}-\ln q_{2}^{0})+\mathbf{o}\left( \frac{1}{q_{1}-q_{2}}\right)
\\ 
=\frac{2\pi i}{q_{1}-q_{2}}\left( \prod_{j\geq 3}\frac{\mathbf{M}%
_{0}-q_{j}^{0}\mathbf{I}_{n}}{q_{d}-q_{j}^{0}}\right) (\mathbf{M}_{0}-q_{d}%
\mathbf{I}_{n})+\mathbf{o}\left( \frac{1}{q_{1}-q_{2}}\right) \\ 
=\frac{2\pi i}{q_{1}-q_{2}}\mathbf{C}_{0}\left[ \left( \prod_{j\geq 3}\frac{%
\mathbf{J}_{0}-q_{j}^{0}\mathbf{I}_{n}}{q_{d}-q_{j}^{0}}\right) (\mathbf{J}%
_{0}-q_{d}\mathbf{I}_{n})\right] \mathbf{C}_{0}^{-1}+\mathbf{o}\left( \frac{1%
}{q_{1}-q_{2}}\right) .%
\end{array}
\label{A7}
\end{equation}%
Next we invoke (\ref{A2}) and observe that 
\begin{equation}
\left( \prod_{j\geq 3}\frac{\mathbf{J}_{0}-q_{j}^{0}\mathbf{I}_{n}}{%
q_{d}-q_{j}^{0}}\right) (\mathbf{J}_{0}-q_{d}\mathbf{I}_{n})=\mathbf{R}%
\oplus \mathbf{0}_{n-2},  \label{A8}
\end{equation}%
which is due to%
\begin{equation}
\begin{array}{c}
\left( \prod_{j\geq 3}\frac{\mathbf{P}-q_{j}^{0}\mathbf{I}_{2}}{%
q_{d}-q_{j}^{0}}\right) (\mathbf{P}-q_{d}\mathbf{I}_{2})=\left[ \prod_{j\geq
3}\left( \mathbf{I}_{2}+\frac{1}{q_{d}-q_{j}^{0}}\mathbf{R}\right) \right] 
\mathbf{R} \\ 
=\left( \mathbf{I}_{2}+\sum_{j=3}^{n}\frac{1}{q_{d}-q_{j}^{0}}\mathbf{R}%
\right) \mathbf{R}=\mathbf{R;} \\ 
\left( \prod_{j\geq 3}\frac{\mathbf{S}-q_{j}^{0}\mathbf{I}_{n-2}}{%
q_{d}-q_{j}^{0}}\right) (\mathbf{S}-q_{d}\mathbf{I}_{n-2})=\mathbf{0}_{n-2}.%
\end{array}
\label{A9}
\end{equation}%
Note that an essential simplification of (\ref{A8}) is a consequence of $%
\mathbf{R}^{2}=\mathbf{0}_{2},$ yielding (\ref{A9}$_{1}$). Finally,
inserting (\ref{A8}) into (\ref{A7}) delivers the sought result (\ref{A6}).
Admitting $\mathbf{A}=\mathbf{0}_{n}$ in (\ref{A6}) would lead to a
contradiction $\mathbf{0}_{n}=\mathbf{C}_{0}^{-1}\mathbf{A}\mathbf{C}_{0}=%
\mathbf{R}_{0}\oplus \mathbf{0}_{n-2}\neq \mathbf{0}_{n}$, hence $\mathbf{A}%
\neq \mathbf{0}_{n}.\ \blacksquare $

Equation (\ref{A6}) shows that the condition (\ref{A5}) leads to divergence
of $\ln \mathbf{M}\left( \omega \right) $ with $q_{1}\left( \omega \right)
\rightarrow q_{2}\left( \omega \right) $ at $\omega \rightarrow \omega _{0}.$
For a unimodular $\mathbf{M,}$ taking (\ref{A6}) with $q_{1}\left( \omega
\right) -q_{2}\left( \omega \right) \approx 2\delta q$ gives $\ln \mathbf{M}%
\left( \omega \right) =\frac{\pi i}{\delta q}\mathbf{A}+\mathbf{o}\left( 
\frac{1}{\delta q}\right) .$ In the case of 2$\times $2 matrices, $\mathbf{%
A=C}_{0}\mathbf{RC}_{0}^{-1}=\mathbf{M}_{0}-q_{d}\mathbf{I}$ and hence (\ref%
{A6}) provides the leading-order term on the right-hand side of (\ref{13}%
).\smallskip

\noindent \textbf{A}{\textbf{2. Low-frequency asymptotics of }}$\ln \mathbf{M%
}$ \textbf{defined over the Brillouin zone} $\left[ 0,2\pi \right]
\smallskip $

Interest in the 'effective' matrix $\mathbf{Q}_{\mathrm{eff}}=i\mathbf{K}=%
\frac{1}{T}\ln \mathbf{M}$ is often confined to the frequency range $\omega
\in \left[ 0,\Omega \right] $ occupied by the first passband, i.e. by the
first Floquet branch. The logarithm of a unimodular $\mathbf{M}$ does not
diverge at $\omega \rightarrow \Omega $ if, contrary to the conventional
definition, its eigenvalues $\ln q$ are defined on the zeroth Riemann sheet
with a cut $\arg q=0,2\pi .$ Like any other $\ln \mathbf{M}$, it is also
continuous for $\omega \rightarrow 0.$ We will, however, demonstrate that
its low-frequency asymptotics has no physical sense and thus the so defined $%
\ln \mathbf{M}\left( \omega \right) $ is of little if any practical value.

For brevity, consider the case of a 2$\times $2 matrix $\mathbf{Q}\left(
y\right) $ given by (\ref{15.0}), in which, however, we keep arbitrary
periodic $\rho \left( y\right) ,~\mu \left( y\right) $ instead of $\rho
_{j},~\mu _{j}$. The matrix $\mathbf{M}\left( T,0\right) \equiv $ $\mathbf{M}
$ expands as the power series 
\begin{equation}
\begin{array}{c}
\mathbf{M=I}+\int_{0}^{T}\mathbf{Q}\left( y\right) \mathrm{d}%
y+\int_{0}^{T}\int_{0}^{y_{1}}\mathbf{Q}\left( y\right) \mathbf{Q}\left(
y_{1}\right) \mathrm{d}y\mathrm{d}y_{1}+... \\ 
=\mathbf{I}+i\omega T\left( 
\begin{array}{cc}
0 & \left\langle \mu ^{-1}\right\rangle \\ 
\left\langle \rho \right\rangle & 0%
\end{array}%
\right) +\frac{1}{2}\left( i\omega T\right) ^{2}\left( 
\begin{array}{cc}
\left\langle \rho \right\rangle \left\langle \mu ^{-1}\right\rangle +\kappa
& 0 \\ 
0 & \left\langle \rho \right\rangle \left\langle \mu ^{-1}\right\rangle
-\kappa%
\end{array}%
\right) +...,%
\end{array}
\label{A10}
\end{equation}%
where $\left\langle \cdot \right\rangle =\int_{0}^{1}\left( \cdot \right) 
\mathrm{d}\varsigma $ and $\kappa =\int_{0}^{1}\int_{0}^{\varsigma }\left[
\rho \left( \varsigma \right) \mu ^{-1}\left( \varsigma _{1}\right) -\mu
^{-1}\left( \varsigma \right) \rho \left( \varsigma _{1}\right) \right] 
\mathrm{d}\varsigma \mathrm{d}\varsigma _{1}.$ For an oblique propagation ($%
k_{x}=\omega s_{x}\neq 0$), $\rho $ should be pre-multiplied by $%
1-s_{x}^{2}\mu /\rho .$ If the period $T$ consists of two homogeneous
layers, then $\left\langle \rho \right\rangle ,$ $\left\langle \mu
^{-1}\right\rangle $ and $\kappa $ reduce to (\ref{N5.1}).

Reserving the notation $\ln \mathbf{M}$ for the conventionally defined
logarithm of $\mathbf{M,}$ introduce another logarithm $\widetilde{\ln }%
\mathbf{M}$ with the aforementioned 'modified' definition, so that%
\begin{equation}
\begin{array}{c}
\ln \mathbf{M=C}\mathrm{diag}\left( \ln q_{1},\ln q_{2}\right) \mathbf{C}%
^{-1}\ \mathrm{with}\ \ln q=\ln |q|+i\arg q,\ -\pi \leq \arg q<\pi ; \\ 
\widetilde{\ln }\mathbf{M=C}\mathrm{diag}\left( \widetilde{\ln }q_{1},%
\widetilde{\ln }q_{2}\right) \mathbf{C}^{-1}\mathrm{\ with\ }\mathbf{\ }%
\widetilde{\ln }q=\ln |q|+i\arg q,\ \ 0\leq \arg q<2\pi ,%
\end{array}
\label{A11}
\end{equation}%
where $q_{1,2}$ are eigenvalues of $\mathbf{M,}$ and $\mathbf{C}$ is a
matrix of eigenvectors of $\mathbf{M}$. Obviously, taking $\exp $ of both $%
\ln \mathbf{M}$ and $\widetilde{\ln }\mathbf{M}$ returns $\mathbf{M.}$
However, these two matrix logarithms are essentially different. Note that
the standard definition used in (\ref{A11})$_{1}$ allows the Taylor series $%
\ln (1+z)=z-\frac{1}{2}z^{2}+...$ for $z\ll 1,$ whereas $\widetilde{\ln }%
\left( 1+z\right) $ used in (\ref{A11})$_{2}$ is not analytical near $z=0$
and hence does not admit the Taylor expansion. This underlies a drastic
disparity between the low-frequency asymptotics of $\ln \mathbf{M}$ and $%
\widetilde{\ln }\mathbf{M}$.

For small $\omega ,$ when $q_{1,2}\left( \omega \right) $ are close to 1, $%
\widetilde{\ln }\mathbf{M}$ and $\ln \mathbf{M}$ are related as follows%
\begin{equation}
\begin{array}{c}
\widetilde{\ln }\mathbf{M=C}\mathrm{diag}\left( \ln q_{1},\widetilde{\ln }%
q_{2}\right) \mathbf{C}^{-1}=\mathbf{C}\mathrm{diag}\left( \ln q_{1},\ln
q_{2}+2\pi i\right) \mathbf{C}^{-1}= \\ 
\ln \mathbf{M+C}\mathrm{diag}\left( 0,2\pi i\right) \mathbf{C}^{-1}=\ln 
\mathbf{M}+\frac{2\pi i}{q_{2}-q_{1}}\left( \mathbf{M}-q_{1}\mathbf{I}%
\right) ,%
\end{array}
\label{A12}
\end{equation}%
where, with reference to (\ref{16}) and (\ref{A10}),%
\begin{equation}
q_{1,2}\left( \omega \right) =1\pm i\omega T\sqrt{\left\langle \rho
\right\rangle \left\langle \mu ^{-1}\right\rangle }+\frac{1}{2}\left(
i\omega T\right) ^{2}\left\langle \rho \right\rangle \left\langle \mu
^{-1}\right\rangle +O\left( \omega ^{3}\right) .  \label{A13}
\end{equation}%
Hence an explicit difference between $\widetilde{\ln }\mathbf{M}$ and $\ln 
\mathbf{M}$ at $\omega \rightarrow 0$ is%
\begin{equation}
\begin{array}{c}
2\pi i\frac{\mathbf{M}-q_{1}\mathbf{I}}{q_{2}-q_{1}}=\frac{2\pi i}{2i\omega T%
\sqrt{\left\langle \rho \right\rangle \left\langle \mu ^{-1}\right\rangle }%
+O(\omega ^{2})}\times \\ 
\left[ i\omega T\left( 
\begin{array}{cc}
-\sqrt{\left\langle \rho \right\rangle \left\langle \mu ^{-1}\right\rangle }
& \left\langle \mu ^{-1}\right\rangle \\ 
\left\langle \rho \right\rangle & -\sqrt{\left\langle \rho \right\rangle
\left\langle \mu ^{-1}\right\rangle }%
\end{array}%
\right) +\frac{1}{2}\left( i\omega T\right) ^{2}\left( 
\begin{array}{cc}
\kappa & 0 \\ 
0 & -\kappa%
\end{array}%
\right) \right] +\mathbf{O}(\omega ^{2}) \\ 
=\pi i\left( 
\begin{array}{cc}
-1 & \sqrt{\frac{\left\langle \mu ^{-1}\right\rangle }{\left\langle \rho
\right\rangle }} \\ 
\sqrt{\frac{\left\langle \rho \right\rangle }{\left\langle \mu
^{-1}\right\rangle }} & -1%
\end{array}%
\right) -\frac{\pi \kappa T}{2\sqrt{\left\langle \rho \right\rangle
\left\langle \mu ^{-1}\right\rangle }}\omega \left( 
\begin{array}{cc}
1 & 0 \\ 
0 & -1%
\end{array}%
\right) +\mathbf{O}(\omega ^{2}).%
\end{array}
\label{A14}
\end{equation}

The low-frequency asymptotics of $\ln \mathbf{M=Q}_{\mathrm{eff}}T$ readily
follows from (\ref{A10}) on the basis of the Taylor series of $\ln (1+z),$
see its example (\ref{N5}). It has a perfectly clear physical meaning, for $%
\mathbf{Q}_{\mathrm{eff}}$ tends to zero when $\omega \rightarrow 0,$ and to
an appropriate matrix $\mathbf{Q}$ of a homogeneous medium when the
inhomogeneity tends to zero (cf. (\ref{N5}) and (\ref{15.0})). As regards $%
\widetilde{\ln }\mathbf{M}$, Eqs. (\ref{A12}) and (\ref{A14}) show that its
discrepancy with $\ln \mathbf{M}$ is non-zero even at $\omega =0.$ Thus,
contrary to $\ln \mathbf{M,}$ the asymptotics of $\widetilde{\ln }\mathbf{M}$
near $\omega =0$ has no physical sense.\smallskip

\noindent \textbf{A}{\textbf{3. Explicit form (\ref{N7})}}$_{2}${\textbf{\
of the matrix }}$\mathbf{A}$\textbf{\ and its properties}$\smallskip $

Consider the matrix $\mathbf{A=M}\left( \omega _{0}\right) -q_{d}\mathbf{I=M}%
\left( \omega _{0}\right) +\mathbf{I}$ defined at the BZ edge, see (\ref%
{N8.0}). Substituting the propagator $\mathbf{M}$ through a bilayered unit
cell given by (\ref{15}) leads to 
\begin{equation}
\mathbf{A}=\frac{1}{Z_{1}Z_{2}}\left( 
\begin{array}{cc}
-\frac{1}{2}\left( Z_{1}^{2}-Z_{2}^{2}\right) \sin \psi _{1}\sin \psi _{2} & 
i\left( {Z_{2}}\sin \psi _{1}\cos \psi _{2}+{Z_{1}}\sin \psi _{2}\cos \psi
_{1}\right) \\ 
{i}Z_{1}Z_{2}\left( {Z_{1}}\sin \psi _{1}\cos \psi _{2}+{Z_{2}}\sin \psi
_{2}\cos \psi _{1}\right) & \frac{1}{2}\left( Z_{1}^{2}-Z_{2}^{2}\right)
\sin \psi _{1}\sin \psi _{2}%
\end{array}%
\right) ,  \label{A15}
\end{equation}%
where $\psi _{j}=\omega _{0}s_{j}d_{j\text{ }}(j=1,2)$ and $\omega _{0}$ is
implicitly determined by Eq. (\ref{17}) or its equivalent (\ref{N1.1}). In
the following, the reference to $\omega =\omega _{0}$ will be understood.
The objective is to manipulate (\ref{A15}) into a form that is transparent.

Introduce the auxiliary notations%
\begin{equation}
Z_{\pm }=Z_{1}\pm Z_{2},\ \psi _{\pm }=\frac{1}{2}(\psi _{1}\pm \psi _{2}),\
a_{\pm }=Z_{\pm }\cos \psi _{\pm },~b_{\pm }=\frac{1}{2}(\sin \psi _{+}\pm
\sin \psi _{-}).  \label{A16}
\end{equation}%
Note the trigonometric identities%
\begin{equation}
\begin{array}{c}
\sin \psi _{1}\sin \psi _{2}=4b_{+}b_{-}=\cos ^{2}\psi _{-}-\cos ^{2}\psi
_{+}; \\ 
{Z_{2}}\sin \psi _{1}\cos \psi _{2}+{Z_{1}}\sin \psi _{2}\cos \psi
_{1}=a_{+}\sin \psi _{+}-a_{-}\sin \psi _{-}; \\ 
\,{Z_{1}}\sin \psi _{1}\cos \psi _{2}+{Z_{2}}\sin \psi _{2}\cos \psi
_{1}=a_{+}\sin \psi _{+}+a_{-}\sin \psi _{-}.%
\end{array}
\label{A17}
\end{equation}%
Next we use Eq. (\ref{N1.1}), which defines two families of the stopband
bounds $\omega _{0}$ given by either $f_{+}=0$ or $f_{-}=0,$ i.e. by either $%
a_{+}=-a_{-}$ or $a_{+}=a_{-}$ (see (\ref{N1})$_{2}$ and (\ref{A16})).
Combining these equations with (\ref{A17}) leads to the following
alternative expressions for the diagonal and off-diagonal elements of $%
\mathbf{A}$: 
\begin{equation}
A_{11}=-A_{22}=-2\frac{Z_{+}Z_{-}}{Z_{1}Z_{2}}\,b_{+}b_{-}=\pm \frac{2}{%
Z_{+}Z_{-}}\,a_{+}a_{-}=\pm (\cos \psi _{1}+\cos \psi _{2}),  \label{A18}
\end{equation}%
and 
\begin{equation}
A_{12}=\frac{i2}{Z_{1}Z_{2}}\,a_{\pm }b_{\pm }=\mp \frac{i2}{Z_{1}Z_{2}}%
\,a_{\mp }b_{\pm },\ A_{21}=i2\,a_{\pm }b_{\mp }=\mp i2\,a_{\mp }b_{\mp },
\label{A19}
\end{equation}%
where%
\begin{equation}
2a_{\mp }b_{\mp }=Z_{1}\sin \psi _{1}-Z_{2}\sin \psi _{2},\ 2a_{\pm }b_{\pm
}=\pm \left( Z_{2}\sin \psi _{1}-Z_{1}\sin \psi _{2}\right) .  \label{A20}
\end{equation}%
Except for the first expression in (\ref{A18}), all others may be called
conditional as they depend on which of the two families of $\omega _{0}$
they are referred to. The compact form of these expressions in (\ref{A18})-(%
\ref{A20}) implies that the upper/lower signs and, simultaneously, the
upper/lower subscripts are related to $f_{+}=0$ and to $f_{-}=0$,
respectively. By using these expressions, Eq. (\ref{A15}) can be recast in
the form%
\begin{equation}
\begin{array}{c}
\mathbf{A=\pm }\left( 
\begin{array}{cc}
\cos \psi _{1}+\cos \psi _{2} & \frac{i}{Z_{1}}\sin \psi _{1}-\frac{i}{Z_{2}}%
\sin \psi _{2} \\ 
iZ_{2}\sin \psi _{2}-iZ_{1}\sin \psi _{1} & -\cos \psi _{1}-\cos \psi _{2}%
\end{array}%
\right) \\ 
=\pm \left( \mathrm{e}^{\mathbf{Q}_{2}d_{2}}\mathbf{G}+\mathbf{G}\mathrm{e}^{%
\mathbf{Q}_{1}d_{1}}\right) ,\ \mathrm{where\ }\mathbf{G}=\left( 
\begin{array}{cc}
1 & 0 \\ 
0 & -1%
\end{array}%
\right) ,%
\end{array}
\label{A21}
\end{equation}%
with $\mathbf{\pm }$ corresponding to $f_{\pm }=0$ as above. This is Eq. (%
\ref{N7})$_{2}$ presented in \S 4.1.2.

By the definition, $\mathbf{A=M}\left( \omega _{0}\right) -q_{d}\mathbf{I}=%
\mathbf{0}$ for a homogeneous medium, in which case (\ref{A21}) holds with $%
Z_{1}=Z_{2},\ s_{1}=s_{2}$ and with $\psi _{1}+\psi _{2}=\pi \left(
2n+1\right) $ due to $\omega =\omega _{0}.$ The matrix $\mathbf{A}$ for a
periodically bilayered medium may incidentally vanish if both $\cos \psi
_{+} $ and $\cos \psi _{-}$ at $\omega =\omega _{0}$ happen to turn to zero
at once, i.e., if $\psi _{1}$ and $\psi _{2}$ in (\ref{A21}) differ by $\pm
\pi $ and in addition one of $\psi _{1,2}$ is equal to $2\pi n$. In general, 
$\mathbf{A}$ is non-semisimple with a zero eigenvalue and hence it must also
admit a dyadic representation via its null vector $\mathbf{u.}$ This
representation further specifies due to the identity $\mathbf{M}^{-1}=%
\mathbf{TM}^{+}\mathbf{T}$ following from (\ref{8}), which may also be
combined with the material-symmetry relation $\mathbf{M}=\mathbf{GM}^{\ast }%
\mathbf{G}$ to give $\mathbf{M}^{-1}=\mathbf{JM}^{\mathrm{T}}\mathbf{J}%
^{-1}, $ where $\mathbf{J=TG;}$ $\mathbf{T}$ is a matrix with zero diagonal
and unit off-diagonal elements; $^{\ast }$ means complex conjugate and $^{+}$
Hermitian adjoint. Hence 
\begin{equation}
\mathbf{A=u}\otimes \mathbf{v}\left( =u_{i}v_{j}\right) \mathbf{,\ }\mathrm{%
where\ }\mathbf{Au=0,\ v=Ju=Tu}^{\ast }\mathbf{,\ }u_{i}v_{i}=0.  \label{A22}
\end{equation}%
Indeed, Eq. (\ref{A21}) may be re-arranged in the form%
\begin{equation}
\mathbf{A}=\frac{\pm 1}{iZ_{1}\sin \psi _{1}-iZ_{2}\sin \psi _{2}}%
\begin{pmatrix}
\cos \psi _{1}+\cos \psi _{2} \\ 
iZ_{2}\sin \psi _{2}-iZ_{1}\sin \psi _{1}%
\end{pmatrix}%
\otimes 
\begin{pmatrix}
iZ_{1}\sin \psi _{1}-iZ_{2}\sin \psi _{2} \\ 
\cos \psi _{1}+\cos \psi _{2}%
\end{pmatrix}%
,  \label{A23}
\end{equation}%
which satisfies (\ref{A22}).


\bigskip

\begin{thebibliography}{99}
\bibitem{Pease} M.C. Pease, III, Methods of Matrix Algebra, Academic Press,
New York (1965).

\bibitem{B} L. Brillouin, Wave Propagation in Periodic Structures, Dover,
New York (1953).

\bibitem{MW} W. Magnus and S. Winkler, Hill's Equation, Interscience, New
York (1966).

\bibitem{N} A.N. Norris, Waves in periodically layered media: A comparison
of two theories, \textit{SIAM J.\ Appl. Math.} \textbf{53} (1993) 1195-1209.

\bibitem{PBG} C. Potel, J.-F. de Belleval and Y. Gargouri, ``Floquet waves
and classical plane waves in an anisotropic periodically multilayered
medium: Application to the validity domain of homogenization" \textit{J. Acoust.
Soc. Am.} 97, 2815-2825 \textbf{(}1995).

\bibitem{WR} L. Wang and S.I. Rokhlin, ``Floquet wave homogenization of
periodic anisotropic media" \textit{J. Acoust. Soc. Am.} 112, 38-45 (2002).

\bibitem{WaMot} A.L. Shuvalov, O. Poncelet and M. Deschamps, ``General
formalism for plane guided waves in transversely inhomogeneous anisotropic
plates"\ \textit{Wave Motion} 40, 413-426 (2004).

\bibitem{WaMot1} A.L. Shuvalov, O. Poncelet and A.P. Kiselev,
\textquotedblleft Shear horizontal waves in transversely inhomogeneous
plates\textquotedblright\ \textit{Wave Motion} 45, 605-615 (2008).

\bibitem{BH} A.M. Braga and G. Herrmann, ``Floquet waves in anisotropic
periodically layered composites"\ \textit{J. Acoust. Soc. Am.} 91, 1211-1227 (1992).

\bibitem{HJ} R.A. Horn and C.R. Johnson, Topics in Matrix Analysis,
Cambridge University Press, Cambridge (1991).

\bibitem{WaMot2} A.L. Shuvalov, ``On the theory of plane inhomogeneous waves
in anisotropic elastic media" \textit{Wave Motion} 34, 401-429 (2001) (Erratum, 
\textit{ibid.} 36, 305 (2002)).

\bibitem{ZBG} F. Zolla, G. Bouchitte and S. Guenneau, ``Pure currents in
foliated waveguides" \textit{Q. J. Mech. Appl. Math.} 61, 453-474 (2008).

\bibitem{ACG} S.M. Adams, R.V. Craster and S. Guenneau, ``Guided and standing
Bloch waves in periodic elastic strips" \textit{Waves Rand. Comp. Med.} 19, 321-346
(2009).
\end{thebibliography}
\end{document}